# Drift-free Fabry-Perot-cavity-based optical refractometry – accurate expressions for assessments of gas refractivity and density

Ove Axner,[1,a)] Isak Silander,[1] Thomas Hausmaninger,[1] and Martin Zelan[2]

[1]*Department of Physics, Umeå University, SE-901 87 Umeå, Sweden*
[2]*Measurement Science and Technology, RISE Research Institutes of Sweden, SE-501 15 Borås, Sweden*

Fabry-Perot (FP) cavity based optical refractometry (OR), FPC-OR, in which the frequency of laser light locked to a longitudinal mode of a FP cavity is monitored while the amount of gas in the cavity is being changed, has repeatedly been shown to have a high potential for characterization of gases, in particular their refractivity and density. Although the precision of this type of techniques can be exceedingly high, the accuracy is often limited by a number of phenomena and non-linear processes, some of which have a technical origin, with the most prominent being drifts of the length of the cavity. Recent advancements in the field, comprising the use of Dual Fabry-Perot cavity OR (DFPC-OR), and in particular, proposed novel methodologies based on drift-free DFPC-OR (DF-DFPC-OR), have alleviated much of these problems, and therefore open up for new levels of accurate characterization of gases. However, to be able to use these techniques for highly accurate assessments, general and explicit expressions are needed for how the gas density, $\rho_n$, is related to the refractivity of the gas, $n-1$, and how the latter depends on the change in frequency of laser light that follows an evacuation of the cavity, $\Delta v_l$. This paper presents such relations that properly acknowledge the influence of all conceivable phenomena and higher order (non-linear) contributions to the assessment of these entities under drift-free conditions, i.e. *i*) the instantaneous deformation of the cavity (spacer and mirrors) due to the presence of the gas; *ii*) the penetration of light into the stacks of coating of the mirrors; *iii*) its dispersion; *iv*) the dispersion of the gas; *v*) the change of the frequency of the cavity mode as the cavity is evacuated; *vi*) the Lorentz-Lorenz equation; *vii*) higher order virial coefficients; and *viii*) the refractivity or density of the residual gas after evacuation, for both the cases with and without relocking of the laser to other cavity modes. The relative contributions of each of the aforementioned phenomena to measurements of gas refractivity and density are individually assessed. It is shown, among other things, that, the influence of cavity deformation is larger for an open cavity (placed in a compartment in which gas is introduced) than a closed one; if not accounted for appropriately, under standard pressure and temperature conditions it can contribute to the accuracy of the technique with up to a few times $10^{-3}$ if an open cavity is used, while it can be an order of magnitude smaller, i.e. in the order of $10^{-4}$, for a well-designed closed cavity. It is also shown that the non-linearities from dispersion are smaller for the cases when relocking takes place than when it is not used. The expressions derived can serve as a basis for assessment of refractivity or gas density, and changes in such, in future realizations of OR in such a way that they can fully benefit from the extraordinary power of DF-DFPC-OR. [Doc. ID XXXXX]

## I. INTRODUCTION

Optical Refractometry (OR) is a technique for assessment of refractivity that repeatedly has proven to be a powerful tool for characterization of gases, primarily their refractivity, density, and pressure.[1-8] It has demonstrated an exceptional high precision, in some cases several orders of magnitude higher than other techniques, for characterization of gases.[4, 5, 9-20] Although it can be used for assessment of both gas density and pressure, there are several advantages of assessing the presence of a gas in terms of density than in pressure, of which one important is that in a finite volume, the former has a much weaker temperature dependence than the latter.

Since OR assesses refractivity, and the first order dependence of density on refractivity is temperature independent,[18-24] gas density can be assessed by OR with virtually no temperature dependence. Examples of situations when this is of importance is when the presence of gas is to be assessed under dynamic conditions or when the temperature is poorly known or controlled, when small gas leaks under high pressure conditions are to be assessed, or when highly accurate reference gas mixtures are to be produced. When OR is performed, the highest precision is obtained by the use of Fabry-Perot (FP) cavities, here referred to as FP-cavity-based OR (FPC-OR), in which the change in the frequency of a laser field that is locked to a longitudinal mode of the cavity is measured as the cavity is filled or evacuated.[1-7,11-20, 24-27]

However, although the first order dependence of density on refractivity does not have any temperature dependence, because of practical reasons, the assessment of the latter by OR is not always fully temperature independent, mainly caused by thermal expansion of parts of the interferometer; in the case of a FP cavity, primarily by the spacer between the mirrors. To minimize these effects, FP cavities are often made by a low thermal expansion material, e.g. Zerodur or an ultra-low expansion (ULE) material.[1, 3, 4, 6, 7, 11, 17-19] Despite this, the length of a FP cavity can still change due to residual thermal expansion, creeping, and various types of relaxations in the cavity spacer material, which thus will adversely affect assessments of both refractivity and gas density.[3, 4, 6, 7, 11, 17, 18]

A means to mitigate this comprises the use of two FP cavities,[2, 3, 6, 7, 13, 18-20] here referred to as a Dual Fabry-Perot Cavity (DFPC). In this case, one cavity is used as the measurement cavity while the other one is the reference cavity. By locking two lasers, each to a longitudinal mode of its own cavity, and mixing the two laser beams on a photo detector to produce a beat signal, it is assumed that the differential drift in length between the two cavities will be reduced, thus minimizing the effects of thermal drifts and relaxations on the assessment. Besides minimizing the effects of common-mode drifts, the introduction of a reference cavity also introduces a necessary reference to which the measured change in frequency of the measurement cavity can be directly assessed. However, despite the

---
a) Author to whom correspondence should be addresses; electronic mail: ove.axner@umu.se

use of DFPC, it has been found that the measurements in general are still significantly affected by, and in many cases even limited by, residual differential drifts of the lengths of the two cavities.[18]

It is clear that the technique would benefit significantly if it could be performed under drift-free conditions, here referred to as DF-DFPC-OS. Although Egan et al. recently have reported on a DFPC-OR system with a minimum of differential relaxations, obtained by long-term baking of a dual-cavity spacer that has been placed in a temperature controlled compartment,[19] we have chosen to alleviate these problems, and to contribute further to the advancement of FPC-OR techniques, by an alternative means. Based on earlier demonstrations of Khélifa et al. as well as Silander et al.,[3,17] using Allan plots, that drifts of a single FP cavity affects the assessment of refractivity predominantly for "long" averaging times, we predict that the influence of drifts can be reduced, and, in fact, even eliminated (i.e. be made insignificant with respect to other sources of uncertainties), not only in single FPC-OR system but also in DFPC-OR system, and even in systems not exposed to elaborate baking procedures, by performing measurements on "short" time scales. In an accompanying paper,[26] we therefore first show that the same holds for DFPC-OR. Based on this, we then propose and present some possible methodologies for DF-DFPC-OR, based on this principle, that can be made by rapidly exchanging gas volumes, referred to as *fast switching* DFPC-OR, henceforth denoted FS-DFPC-OR. In a third work,[27] finally, we estimate the residual precision, accuracy, and temperature dependence of the FS-DFPC-OR methodologies that are portrayed in Ref. [26].

However, to benefit fully from this new concept, in particular to allow for high accuracy assessments of refractivity and gas density, general and accurate expressions are needed for how a measured change in the frequency of the laser light that follows a cavity evacuation is related to the refractivity and the density of a gas. Although various versions of such expressions previously have been derived and discussed in a number of works,[4-6, 9, 11, 12, 14-17, 25] they are not always valid with sufficient accuracy when highly accurate assessments are to be performed. The reason is that there are a number of phenomena (some of which have technical origin) and non-linear dependencies that are not always appropriately taken into account; for the assessment of refractivity the continuous change of the frequency of the cavity mode as the cavity is evacuated, the instantaneous deformation of the cavity due to the presence of the gas, the penetration of light into the coating of the mirrors, including its dispersion, the dispersion of the gas, and the influence of residual gas density in the cavity following an evacuation. For the assessment of gas density additionally also the Lorentz-Lorenz equation and higher order virial coefficients are also not always appropriately taken into account. If not all these are accounted for properly, FPC-OR (and thereby DFPC-OR) will inexorably experience a limited accuracy. Under standard pressure and temperature (STP) conditions, such systematic errors can be in the $10^{-4}$ to $10^{-3}$ range (depending on the type of cavity used), which is insufficient for many types of applications.

In order to remedy this, this paper presents a derivation of proper expressions for first the refractivity and then the density of the gas in a cavity in terms of the measured change in frequency of a laser field locked to a longitudinal mode of the cavity as the cavity is evacuated that is valid under drift-free conditions, i.e. when slow processes, primarily those caused by thermal drifts and relaxations of the spacer material, play a negligible role to the assessments. This is done by not only considering the leading terms in the expressions for how the refractivity, $n-1$, depends on the measured change in laser light frequency, $\Delta v_l$, and how the density of molecules, $\rho_n$, is related to the refractivity, $n-1$, but also acknowledge the influence of the aforementioned phenomena and higher order (or non-linear) contributions to these relations. The relative contributions of these effects are scrutinized in a systematic way under some typical conditions and the combined effects of all processes are presented in a coherent description of FPC-OR that, to the authors' knowledge, has not previously been given in the literature.

In addition, the paper deals with both the cases when the laser light in the measurement cavity is locked to the same cavity mode during the entire evacuation of the cavity (henceforth referred to as without relocking") and when it is relocked to another cavity mode. It is shown that the non-linearities affect these two modes of detection in dissimilar manners.

Before the expanded expressions for the refractivity and the density of molecules are derived, which to a large extent is presented in the supplementary material, the basis of OR for assessment of refractivity and gas density in terms of the frequency shift of a laser locked to a mode of an ideal optical cavity containing a thin ideal gas as the cavity is evacuated is shortly given. These expressions then serve as a basis for the more extended descriptions, derived thereafter.

By this, the present paper lays the foundation for a variety of future realizations of DF-FPC-OR and DF-DFPC-OR in which the frequency(ies) of (a) narrow-linewidth laser(s) is (are) locked to (an) individual mode(s) of the cavity. Moreover, it does so in such a way that these types of technique can be used for highly accurate assessments of refractivity and gas density under a variety of conditions, in particular the important and commonly occurring non-temperature stabilized ones. It is hoped that the three papers, the present and the Refs. [26, 27], can serve as a basis for future design, construction, and realization of non-complicated and versatile FPC-OR and FS-DFPC-OR instrumentation, and measurement procedures in such, for assessment of refractivity and gas density (including changes in such) that can be used under a variety of conditions and benefit as much as possible from the extraordinary potential power of FPC-OR.

## II. THE BASIS OF OPTICAL REFRACTOMETRY FOR ASSESSMENT OF REFRACTIVITY AND GAS DENSITY

When deriving general expressions for the refractivity and the density of a gas in terms of the shift of the frequency of a laser locked to a mode of an optical cavity as the cavity is evacuated, it is advantageous to have access to the corresponding first order expressions since these can serve as the basis for the more general expressions. Such expression can be obtained by considering a thin ideal gas and an ideal optical cavity. Although such have been given repeatedly in the literature, in various forms, see e.g. Refs. [3-7, 9-18, 24, 25, 28-30], for convenience, and to establish the nomenclature used in this work, they will be reproduced here, both for the cases without and with relocking.

### A. A first order expression for gas refractivity in terms of the shift of the frequency of a mode of an ideal optical cavity as the cavity is evacuated

As is alluded to in the supplementary material (part 1), it is often assumed that the frequency, $v_q$, of the $q$:th longitudinal mode of a Fabry-Perot cavity containing a gas with an index of refraction of $n$ can be expressed as [3-5, 10, 11, 13-15, 17, 18, 24, 28-30]

$$v_q(n) = \frac{qc}{2nL_0}, \tag{1}$$

where $c$ is the speed of light in vacuum and $L_0$ the length of the empty cavity. It is also often assumed that when such a cavity is being evacuated, the frequency of the cavity mode, whose frequency nominally is $\nu_0$, is shifted an amount $\Delta\nu_{cm}$ that is directly proportional to refractivity, i.e. given by

$$\frac{\Delta\nu_{cm}}{\nu_0} = n - 1. \tag{2}$$

When refractivity is to be assessed, and for the case when the frequency of the laser is locked to the same cavity mode during the entire gas evacuation process, it is therefore often assumed that it is directly given by the shift of the frequency of the laser light, $\Delta\nu_l$, defined as $\nu_l(n=1) - \nu_l(n)$,[31] by

$$n - 1 = \frac{\Delta\nu_l}{\nu_0}. \tag{3}$$

However, since the shift of the *q*:th longitudinal mode of the cavity can be large (for gas under STP conditions several tens of GHz), in many cases larger than what a narrow linewidth laser can be rapidly scanned, it is possible, as an alternative, to make a controlled mode jump, i.e. to relock the laser to a different cavity mode. When this takes place, the shift of the frequency of the laser field is given by

$$\Delta\nu_l = \Delta\nu_{cm} - \Delta q \cdot \nu_{FSR}^0, \tag{4}$$

where $\Delta q$ represents the number of modes by which the laser field is shifted during the relocking process and $\nu_{FSR}^0$ is the free spectral range of the empty cavity, given by $c/(2L_0)$. This implies that it is often assumed that the refractivity can be assessed from a measurement of $\Delta\nu_l$ and $\Delta q$ by

$$n - 1 = \frac{\Delta\nu_l}{\nu_0} + \Delta q \cdot \frac{\nu_{FSR}^0}{\nu_0} = \frac{\Delta\nu_l}{\nu_0} + \frac{\Delta q}{q_0}, \tag{5}$$

where $q_0$ denotes the number of the cavity mode that the laser addresses in an empty cavity (which is given by $\nu_0 / \nu_{FSR}^0$).

However, as can be seen from Eq. (S2) in the supplementary material (part 1), Eq. (2), and thereby also the Eqs. (3) and (5), are in reality only the first order terms in series expansions of more general expressions that originate from the fact that the cavity mode is continuously shifted as the cavity is evacuated. Hence, the Eqs. (3) and (5) are not fully accurate, not even when a thin, ideal gas is considered in an ideal optical cavity.

In addition, there are also a number of technical reasons for why these expressions are not fully correct. As is discussed below, as well as in part 2 of the supplementary material, the major one is that the cavity spacer is exposed to thermal drifts and relaxations so its length changes in an uncontrolled manner. Others are that: *i*) the spacer and the mirrors are being deformed by the gas in the cavity; *ii*) the standing-wave light field inside the cavity penetrates into the mirror coatings, providing a finite penetration depth that, in turn, has a wavelength dependence, normally referred to as mirror dispersion; and *iii*) the index of refraction of a gas has dispersion. Despite all this, the Eqs. (3) and (5) are commonly used for FPC-OR. When this is being done, the accuracy of the technique is significantly limited.

### B. An expression for the density of a thin ideal gas in terms of refractivity

When gas density is to be assessed, it is common to assume that the density of a gas, $\rho_n$, can be assessed from a measurement of its refractivity, $n-1$, according to [4,9,11,14,15,17,25]

$$\rho_n = \frac{2}{3A}(n-1), \tag{6}$$

where $A$ is the (number or molar) polarizability (or refractivity), which is given by the (dipolar) polarizability, $\alpha$, and the diamagnetic susceptibility, $\chi$, that, in CGS units,[32] can be written as [23,33]

$$A = \frac{4\pi}{3}(\alpha + \chi). \tag{7}$$

However, Eq. (6) is solely valid for a thin (in reality an infinitely thin) ideal gas. As is shown by Eq. (S2) in the supplementary material, when the gas cannot be considered to be infinitely thin, the $2(n-1)/3$ entity in Eq. (6) constitutes only the first order term of a series expansion of a more general expression, the Lorentz-Lorenz equation.[29,30,34] In addition, as is shown below, since the molecules in a gas in reality interact with each other, by two-and three-body interactions, it is not always appropriate to consider the gas to be ideal. In such cases it is not sufficient to express gas density in terms of a single term with a linear dependence on $n-1$. All this implies that also Eq. (6) should be used with great caution.

### C. A first order expression for the density of a gas in terms of the shift of the frequency of a laser locked to a mode of an optical cavity as the cavity is evacuated

Based on the Eqs. (3), (5), and (6), it is often tempting to assume that $\rho_n$ can be assessed from an FPC-OR measurement as the cavity is evacuated by expressions that, for the cases without and with relocking, often are written as

$$\rho_n = \frac{2}{3A} \frac{\Delta\nu_l}{\nu_0} \tag{8}$$

and

$$\rho_n = \frac{2}{3A}\left(\frac{\Delta\nu_l}{\nu_0} + \Delta q \cdot \frac{\nu_{FSR}^0}{\nu_0}\right) = \frac{2}{3A}\left(\frac{\Delta\nu_l}{\nu_0} + \frac{\Delta q}{q_0}\right), \tag{9}$$

respectively. Hence, although these expressions look handy and sometimes appear in the literature,[15-17] since they are derived from expressions that are not always fully adequate when real gases are considered; they should only be used with caution, and, in particular, not be used when high accuracy FPC-OR assessments of real gases are to be made.

## III. ASSESSMENTS OF REFRACTIVITY BY DRIFT-FREE FABRY-PEROT-BASED OPTICAL REFRACTOMETRY

As was mentioned in the introduction, when (D)FPC-OR is performed under conditions under which slow drift processes, primarily those caused by thermal drifts and relaxations of the spacer material, play a minor role to the overall uncertainty, referred to as drift-free (DF) conditions, the main causes of uncertainty of an assessment of refractivity have been eliminated. Despite this, it is not trivial to assess the refractivity of a gas from an DF-(D)FPC-OR measurement with high accuracy. The reason is that there are a multitude of other causes why neither the simplest expression for refractivity in terms of shift of the cavity mode, e.g. Eq. (3), nor its non-contracted counterpart, Eq. (S2) in the supplementary material (part 1), are fully correct. Accurate assessments of refractivity can only be performed as long as all processes that influence a measurement have been identified and properly characterized. This has until now been one important impediment for highly accurate assessments of refractivity by (D)FPC-OR.

### A. General expression for the frequency of the *q*:th longitudinal mode of a Fabry-Perot cavity containing gas

One reason why the Eqs. (3) and (7) are not fully adequate is that they are based upon the assumption that the physical length of the cavity is constant, in particular independent of the gas presence. However, since the pressure in the cavity will deform both the spacer and the mirrors (primarily by elongation and bending, respectively), it is not appropriate to consider the cavity length to be a constant; instead, it is in general a function of pressure of the gas.

Another reason is that the field of the standing-wave light inside the cavity penetrates the mirror coating. Because of this, the effective length of the cavity (experienced by the light) will differ from its physical length (over which gas is distributed). Moreover, since the reflectivity of coated mirrors, and thereby the penetration depth, originate from interference of the light between the various layers of the mirror coating, and as such depends on the wavelength of the light, the penetration depth will have a frequency dependence, commonly termed mirror dispersion. It should also be noted that, in general, the index of refraction of the gas has a certain frequency dependence, here referred to as dispersion of the gas. All this implies that the optical length of the cavity in reality has both a pressure and a frequency dependence that potentially can affect an assessment.

The length of the spacer has also a dependence on temperature. However, since optical cavities in general are made of low-thermal expansion materials, and this work focusses on DF-(D)FPC-OR, which addresses assessments under drift-free conditions, as is discussed in more detail in one of our accompanying works,[27] this effect has in general solely a limited influence of the performance of the technique.

As is scrutinized in some detail in part 2 of the supplementary material, the frequency of the $q$:th cavity mode, instead of being given by Eq. (1), should in reality be given by[35]

$$\nu_q(n) = \frac{qc}{2\left[n(\nu_q)L(p) + 2L_{pd}(\nu_q)\right]}, \quad (10)$$

where $L(p)$ is the physical length of the cavity, defined as the distance between the surfaces of the mirrors (which thus represents the length over which the cavity can be filled with gas), given by $L_0 + \Delta L_E(p)$, where $L_0$ is the physical length of the empty cavity while $\Delta L_E(p)$ is the elongation of the cavity length by deformation due to the presence of the gas. $L_{pd}(\nu_q)$ is the penetration depth of the light into the coating of a mirror, given by $L_{pd}^0 + \Delta L_{pd}(\nu)$ where $L_{pd}^0$ is the frequency independent part, evaluated at a given frequency, while $\Delta L_{pd}(\nu)$ represents the frequency dependent part of the penetration depth, i.e. the mirror dispersion. Each of these phenomena are discussed in more detail in the supplementary material (part 2).

**B. General expression for the frequency shift of a laser field locked to a mode of a drift-free optical cavity as the cavity is evacuated**

Part 2 of the supplementary material provides, based on Eq. (10), a derivation of a general (although extensive) expression for the frequency shift of the laser field as the cavity is evacuated, $\Delta\nu_l(n_i, n_f, \Delta q)$, given by Eq. (S41), in terms of the index of refraction of the gas both prior to and after an evacuation, denoted $n_i$ and $n_f$, that take cavity deformation, the finite penetration depth of light into the coatings of the mirrors, its dispersion, and the dispersion of the gas into account. The subscripts $i$ and $f$ stand for the initial and the final conditions, respectively. In the case relocking takes place, Eq. (S41) is also expressed in terms of the number of modes the laser field is shifted, $\Delta q$. After an assessment of the relative magnitude of the various contributions of these phenomena to the frequency shift, it was found that this expression could be simplified by series expanding it into a number of small entities. By doing so, and then neglecting all terms that are expected to influence the assessed frequency shift by less than $1:10^{10}$ under STP conditions, a more manageable expression for the frequency shift, expressed in terms of the indices of refraction of the gas both prior to and after evacuation, could be formulated [given by Eq. (S42)] that reads

$$\Delta\nu_l(n_i, n_f, \Delta q) \approx \nu_0 \left\{ \frac{n_i - n_f}{n_i n_f}(1+\varepsilon) - (2\varsigma + \eta)\frac{\Delta\nu_l}{\nu_0} \right.$$
$$\left. - \frac{\Delta q}{q_0}\frac{1}{n_i}\left[1 - \varepsilon(n_i - 1) - (2\varsigma + \eta)\frac{\Delta\nu_l}{\nu_0}\right] \right\}. \quad (11)$$

Here, $\varepsilon$ is a parameter that represents the effect of an instantaneous deformation of the spacer material and the mirrors (effectively a change in cavity length) caused by the pressure of the gas in the cavity, given by Eq. (S21), here referred to as the cavity deformation parameter, while $\varsigma$ and $\eta$ characterize the dispersion of the mirrors and the gas, given by the Eqs. (S32) and (S36), respectively (where $2\varsigma$ corresponds to $\varepsilon_\alpha$ in Ref. [6]). All entities are derived and fully defined in part 2 of the supplementary material.

Equation (11) shows that for the case with no relocking, the parameter that represents the instantaneous change in length of the cavity due to the presence of the gas, represented by $\varepsilon$, contributes to the shift of the laser frequency by an amount that is proportional the difference in refractivity, $n_i - n_f$, while dispersion (from the mirrors as well as the gas, i.e. $\varsigma$ and $\eta$) influences the shift in proportion to the relative shift of the laser frequency, i.e. by $\Delta\nu_l/\nu_0$. Since, for the case with no relocking and according to the Eqs. (3) and (6), both these entities represent the amount of gas in the cavity, which implies that they are identical. Hence, this shows that the elongation and dispersion will affect an assessment in similar manners, i.e. $\Delta\nu_l(n_i, n_f, \Delta q = 0)$ will mainly be given by $\nu_0(n_i - n_f)(1 + \varepsilon - 2\varsigma - \eta)/n_i n_f$.

When relocking takes place, the relative shift of the frequency of the laser, $\Delta\nu_l/\nu_0$, is significantly smaller than $n_i - n_f$. Under these conditions, the shift of the laser frequency is significantly less affected by dispersion than by the deformation of the cavity.

As is discussed both in the supplementary material (part 2) and in our accompanying work,[27] it is estimated that for a well-designed closed cavity, $\varepsilon$ can take values around $10^{-4}$, although it can be an order of magnitude larger for an open cavity (placed in a compartment in which gas is introduced).[36] Moreover, $\varsigma$ can range from $10^{-6}$ to $10^{-3}$ (depending on the dispersion of the mirrors) while $\eta$ typically takes a value around $10^{-6}$ (for the case with nitrogen at 1530 nm, $1.2 \times 10^{-6}$).

**C. General expression for the refractivity in terms of the frequency shift of the laser locked to a mode of a drift-free optical cavity as the cavity is evacuated**

Since one of the main aims of FPC-OR is to assess refractivity by the use of a measured frequency shift, Eq. (11), which provides an expression for the shift of the laser frequency in terms of various entities, among them the index of refraction of the gas both prior to and after evacuation, is not formulated in an optimal manner. Rewriting Eq. (11) in terms of the refractivity of the gas prior to and after the evacuation, i.e. in terms of $n_i - 1$ and $n_f - 1$, respectively, as is done in Eq. (S43) in the supplementary material (part 2), results in an expression that can be solved for the refractivity of the gas in the cavity prior to the evacuation, $n_i - 1$, given by Eq. (S44), that reads

$$n_i - 1 =$$

$$\frac{\frac{\Delta q}{q_0} + \Upsilon(\varsigma,\eta)\frac{\Delta\nu_l}{\nu_0} + 2\varsigma\frac{\Delta\nu_l}{\nu_0}\frac{\Delta q}{q_0} + (n_f - 1)\left[1 + \varepsilon + \Upsilon(\varsigma,\eta)\frac{\Delta\nu_l}{\nu_0} + \frac{\Delta q}{q_0}\right]}{1 + \left(1 + \frac{\Delta q}{q_0}\right)\varepsilon - \Upsilon(\varsigma,\eta)\frac{\Delta\nu_l}{\nu_0} - (n_f - 1)\left[\Upsilon(\varsigma,\eta)\frac{\Delta\nu_l}{\nu_0} - \varepsilon\frac{\Delta q}{q_0}\right]}$$
, (12)

where $\nu_0$ now corresponds to the frequency of laser lights locked to the $q$:th longitudinal mode of an empty cavity and where $\Upsilon(\varsigma,\eta)$ is given by $1 + 2\varsigma + \eta$. This expression can in principle be used for assessment of refractivity from a measurement of $\Delta\nu_l$ and $\Delta q$ if all the other entities are known.

However, Eq. (12) shows that $n_i - 1$ is neither linear in $\Delta\nu_l/\nu_0$, nor in $\Delta q/q_0$. This makes if rather complex to analyze; it is, for example, not directly evident what information it can provide to designers or users of FPC-OR instrumentation. However, since all non-unity entities in this expression are significantly smaller than unity, it is possible to estimate their (non-linear) influences on an assessment of refractivity by series expanding it. Such an expression can be made arbitrarily accurate by the inclusion of an appropriate number of terms in the expansion. Although different instrumentation, realizations, and assignments provide different conditions, it is by this means possible to scrutinize it for a few typical situations. We will do so here and specifically distinguish between the cases when the laser field is locked to a given cavity mode during the entire evaluation process (i.e. for no relocking, when $\Delta q = 0$) and when it is relocked to a different cavity mode.

**D. Simplified expression for the refractivity in terms of the frequency shift of the laser locked to a mode of a drift-free optical cavity as the cavity is evacuated with no relocking – Influence of higher order terms, cavity deformation, dispersion, and residual gas**

The shift of a mode of a cavity that originally is filled with gas and then rapidly evacuated, $\Delta\nu_{cm}$, is, according to Eq. (2), roughly given by $(n-1)\nu_0$. This implies that, for the case with no relocking, for which $\Delta\nu_l = \Delta\nu_{cm}$, $\Delta\nu_l/\nu_0$ is basically given by $n-1$, which for a gas under STP conditions takes a value of a few times $10^{-4}$ ($3 \times 10^{-4}$ for N2, henceforth referred to as $N_2^{STP}$, at 1.5 μm). In addition, as was alluded to above, for a well-designed cavity, $\varepsilon$ can usually take values of similar magnitude, $\varsigma$ can range between $10^{-6}$ to $10^{-3}$, while $\eta$ typically takes a value of $10^{-6}$. It can moreover be concluded that although the cavity can be pumped down to any pressure, whereby $n_f - 1$, representing the refractivity of the residual amount of gas, henceforth referred to as the residual refractivity, can take any value below that of $n_i - 1$, it is possible with modern gas systems to evacuate the cavity so that the residual gas refractivity is a minor fraction of $n_i - 1$; it suffices to evacuate the cavity down to 1 mTorr to produce a final refractivity, $n_f - 1$, of a few times $10^{-10}$ ($4 \times 10^{-10}$ for $N_2$ at 1.5 μm). All this implies that, as long as the cavity is evacuated to a fraction of its initial pressure, Eq. (12) can be series expanded in terms of a number of entities, viz. $\varepsilon$, $\varsigma$, $\eta$, $n_f - 1$, $\Delta\nu_l/\nu_0$, and $\Delta q/q_0$.

Extracting the leading $(\Delta\nu_l/\nu_0)$ term of such an expression, and neglecting terms that are $10^{-10}$ times smaller than this under STP conditions, implies, as is shown by Eq. (S45), that the expression for $n_i - 1$ given above, Eq. (12), can be expressed as[37]

$$n_i - 1 = \Omega(\varepsilon,\varsigma,\eta)\frac{\Delta\nu_l}{\nu_0}\left[1 + \Omega(\varepsilon,\varsigma,\eta)\frac{\Delta\nu_l}{\nu_0} + \left(\frac{\Delta\nu_l}{\nu_0}\right)^2\right]$$
$$+ (n_f - 1)\left[1 + 2\Omega(\varepsilon,\varsigma,\eta)\frac{\Delta\nu_l}{\nu_0} + 3\left(\frac{\Delta\nu_l}{\nu_0}\right)^2\right]. \quad (13)$$

where we have used $\Omega(\varepsilon,\varsigma,\eta)$ as a shorthand notation $1 - \varepsilon + 2\varsigma + \eta + \varepsilon^2$.

This expression shows first of all to what extent the assessment of refractivity is affected by the instantaneous deformation of the cavity due to the presence of gas in the cavity, given by the cavity deformation parameter, $\varepsilon$, as well as dispersion, expressed by $\varsigma$ and $\eta$. It is of importance to conclude that they do not primarily give rise to any non-linear response to the system; they affect an assessment predominantly by a shift in the linear response. This implies that $n_i - 1$ is not primarily solely given by its first order expression, $\Delta\nu_l/\nu_0$, but rather by $\Omega(\varepsilon,\varsigma,\eta)(\Delta\nu_l/\nu_0)$. As is further alluded to in our accompanying paper,[27] this implies that the instantaneous deformation of the cavity due to the presence of gas in the cavity and the dispersion of the gas and the mirrors predominantly affect the accuracy of the instrumentation, but do not predominantly contribute to any non-linear response of the system. Being such, their combined effect can therefore be taken into account by a carefully performed characterization procedure.

Equation (13) then also provides an assessment of the non-linearity in $\Delta\nu_l$. Since $\Delta\nu_l/\nu_0$ is around $3 \times 10^{-4}$ under STP conditions, this implies that the first order non-linearity in $\Delta\nu_l/\nu_0$ needs to be taken into account when a relative accuracy better than a few times $10^{-4}$ is required, while the second order term needs to be taken into account only when higher accuracies (better than $10^{-7}$) are needed. When less dense gases are characterized, the non-linarites plays a successively smaller role. When gases with densities above those at STP conditions are to be addresses, on the other hand, for which the refractivity is higher, the higher order terms play a more significant role

The $n_f - 1$ term indicates simply that the refractivity assessed by a measurement of a shift of a laser frequency in reality is the difference between the refractivity of the gas prior to evacuation and that after, i.e. $(n_i - 1) - (n_f - 1)$. As is further alluded to in one of our accompanying papers, [27] this is of importance when refractivity is to be assessed with high accuracy by these types of techniques.

All this shows that whenever accurate assessments of gas density are to be performed by DF-FPC-OR with no relocking, $n_i - 1$ needs to be expressed beyond the linear relationship that is given by Eq. (3), viz. in terms of $\varepsilon$, $\varsigma$, $\eta$, and $n_f - 1$ as well as higher order terms of $\Delta\nu_l/\nu_0$, preferably by Eq. (13).

**E. Simplified expression for the refractivity in terms of the frequency shift of the laser locked to a mode of a drift-free optical cavity as the cavity is evacuated in the presence of relocking – Influence of higher order terms, cavity deformation, dispersion, and residual gas**

For the case when relocking takes place, it is customary to relock the laser to a mode so that $\Delta\nu_l/\nu_0$ is smaller than $\nu_{FSR}^0/\nu_0$ and thereby also smaller than $\Delta q/q_0$, which instead often is in the order of $n_i - 1$. As also is shown in the supplementary material [Eq. (S66)], this implies that Eq. (12) can be simplified by a series expansion (again neglecting terms that are below $10^{-10}$ of the leading one) to

$$n_i - 1 = \frac{\Delta q}{q}\left[1 - \varepsilon\left(1 + \frac{\Delta q}{q}\right) + \varepsilon^2\right]$$
$$+ \frac{\Delta\nu}{\nu_0}\left(1 + \frac{\Delta q}{q}\right)\left[\Omega(\varepsilon,\varsigma,\eta) - 2\varepsilon(1 - 2\varepsilon)\frac{\Delta q}{q}\right]$$
$$+ \left(\frac{\Delta\nu}{\nu_0}\right)^2\left(1 + \frac{\Delta q}{q}\right)[2\Omega(\varepsilon,\varsigma,\eta) - 1] \quad (14)$$
$$+ (n_f - 1)\left(1 + \frac{\Delta q}{q}\right)\left[1 - 2\varepsilon\frac{\Delta q}{q} + 2\Omega(\varepsilon,\varsigma,\eta)\frac{\Delta\nu_l}{\nu_0}\right].$$

This shows that also when relocking takes place refractivity is non-linear with $\Delta\nu_l/\nu_0$. However, since, in this case, $\Delta\nu_l/\nu_0$ is substantially smaller than for the case when no relocking takes place, this non-linearity is smaller. Also the non-linear dependence on $\Delta q/q_0$ is smaller; this is evidenced by the fact that the first non-linear response with respect to $\Delta q/q_0$ consists of a combined $\varepsilon(\Delta q/q_0)$ term. However, the fact that there are several combined terms, e.g. one containing $(\Delta\nu_l/\nu_0)(\Delta q/q_0)$, makes the situation slightly more difficult to scrutinize. Irrespective of this, it can be concluded that the total non-linearity of the refractivity for the case when relocking takes place is smaller than in the case when no relocking takes place.

## IV. GAS DENSITY IN TERMS OF REFRACTIVITY FOR A REAL GAS

When FPC-OR is used for assessments of gas density, the latter needs to be related to refractivity in an appropriate manner.

### A. Gas density in terms of refractivity

For a real gas, exposed to both two- and three-body interactions, it is customary to relate the index of refraction to the density of molecules according to

$$\frac{n^2-1}{n^2+2} = A_R \rho_n + B_R \rho_n^2 + C_R \rho_n^3 + \ldots, \quad (15)$$

where $A_R$, $B_R$, and $C_R$, are so called virial coefficients.[10,21,28,29,38,39] Since the expression on the left hand side usually is termed the Lorentz-Lorenz equation, while that on the right side represent a virial expansion of the density dependence, this expression will henceforth be referred to as the virial-expanded Lorentz-Lorenz equation.

However, when OR is used for assessment of gas density, this expression is not written in its most suitable form. Since Eq. (15) can be seen as a convergent series of successively higher order terms it is shown in the supplementary material (part 3) that it is possible to express the density as a convergent series of successively higher order terms of the refractivity as

$$\rho_n = A_\rho (n-1) + B_\rho (n-1)^2 + C_\rho (n-1)^3 + \ldots, \quad (16)$$

where the various expansion coefficients, $A_\rho$, $B_\rho$, and $C_\rho$, are explicitly given in Table 1, expressed in terms of the $A_R$, $B_R$, and $C_R$ coefficients.

Table 1. Expressions for the various coefficients used for assessment of gas density in terms of refractivity by use of the Eq. (16) or (17).*

| Coefficient | Relation to the $A_R$, $B_R$, and $C_R$ coefficients |
|---|---|
| $A_\rho$ | $\dfrac{2}{3A_R}$ |
| $B_\rho$ | $-\dfrac{1}{9A_R}\left(1 + 4\dfrac{B_R}{A_R^2}\right)$ |
| $C_\rho$ | $-\dfrac{4}{27}\dfrac{1}{A_R}\left(1 + 2\dfrac{C_R}{A_R^3} - \dfrac{B_R}{A_R^2} - 4\dfrac{B_R^2}{A_R^4}\right)$ |
| $\tilde{B}_\rho$ | $-\dfrac{1}{6}\left(1 + 4\dfrac{B_R}{A_R^2}\right)$ |
| $\tilde{C}_\rho$ | $-\dfrac{2}{9}\left(1 + 2\dfrac{C_R}{A_R^3} - \dfrac{B_R}{A_R^2} - 4\dfrac{B_R^2}{A_R^4}\right)$ |

* For $N_2$ at 632 nm, the $A_R$, $B_R$, and $C_R$ coefficients have values of 4.45 × 10⁻⁶ m³ mol⁻¹, 0.64 × 10⁻¹² m⁶ mol⁻², and 107 × 10⁻¹⁸ m⁹ mol⁻⁹.[29] This implies that under these conditions $B_R/A_R^2$ = 0.032, $C_R/A_R^3$ = 1.2, and $B_R^2/A_R^4$ = 0.0010, which, in turn, implies that the $\tilde{B}_\rho$ and $\tilde{C}_\rho$ coefficients take values of -0.17 and -0.75, respectively.

It is though often more convenient to instead write the expression for the density as

$$\rho_n = \frac{2}{3A_R}(n-1)\left[1 + \tilde{B}_\rho(n-1) + \tilde{C}_\rho(n-1)^2 + \ldots\right], \quad (17)$$

where the leading term, which provides the linear $(n-1)$-dependence, has been extracted and its coefficient, $A_\rho$, explicitly expressed, $2/(3A_R)$. The normalized higher order coefficients, $\tilde{B}_\rho$ and $\tilde{C}_\rho$, which are given by the ratios $B_\rho/A_\rho$ and $C_\rho/A_\rho$, respectively, are likewise given in Table 1, expressed in terms of the $A_R$, $B_R$, and $C_R$ coefficients.

A comparison between the Eqs (6) and (16) shows that, as long as $A_R$ is interpreted as the (number or molar) polarizability for an ideal gas, i.e. $A$, as normally is the case, the leading term in the expression for the real gas corresponds to that for an ideal gas. Equation (17) shows that the other terms, containing $\tilde{B}_\rho$ and $\tilde{C}_\rho$, henceforth simply referred to as higher order terms, represent the "corrections" to the description for an ideal gas. Note that the normalized $\tilde{B}_\rho$ and $\tilde{C}_\rho$ coefficients do not only comprise the higher order virial coefficients, $B_R$, and $C_R$, they also have contributions from the Lorentz-Lorenz equation, which is non-linear in refractivity.

### B. Influence of the higher order terms on the accuracy of the assessment of gas density by measurement of the refractivity

Some typical values of the coefficients of the virial-expanded Lorentz-Lorenz equation (i.e. the $A_R$, $B_R$, and $C_R$ coefficients), the expansion coefficients of the corresponding expressions for the gas density ($A_\rho$, $B_\rho$, and $C_\rho$), as well as the normalized higher order coefficients ($\tilde{B}_\rho$ and $\tilde{C}_\rho$) are estimated for the case with $N_2$ at 632 nm in the footnote of Table 1. Since this shows that $\tilde{B}_\rho$ and $\tilde{C}_\rho$ take values of -0.17 and -0.75 under these conditions, it can be assumed that typical values for these coefficients at other wavelengths or for normal air are similar, i.e. roughly in the order of 10⁻¹ and 10⁰, respectively.[40] Again, since $n-1$ often is around a few times 10⁻⁴ under STP conditions, this shows that it is in general not appropriate to neglect the two non-linear terms in Eq. (17) when highly accurate assessments of gas density are to be done. For $N_2^{STP}$ these terms contribute to the assessment of density roughly by relative values of 5 × 10⁻⁵ and 7 × 10⁻⁸, respectively. This implies that the first of the higher order terms in Eq. (17) needs to be taken into account whenever the density is to be assessed with an accuracy that is better than a few times 10⁻⁵, while the second one has to be included if the accuracy is to be determined with a value that is a couple of orders of magnitude better than this, i.e. in the 10⁻⁷ - 10⁻⁸ range or better.

If gas densities below those at STP are to be assessed, $n-1$ becomes smaller, whereby the higher order terms play a successively smaller role. On the contrary, when gases with densities above those at STP conditions are to be addresses, for which the refractivity is higher, the higher order terms play a more significant role.

## V. ASSESSMENTS OF GAS DENSITY BY DRIFT-FREE FABRY-PEROT-BASED OPTICAL REFRACTOMETRY

By access to explicit expressions for how the gas density depends on refractivity, given by Eq. (17), and how refractivity, in turn, depends on the frequency shift of a laser locked to a mode of a drift-free optical cavity as the cavity is evacuated, given by either of the Eqs. (13) or (14), it is possible to provide expressions for how the density of gas in the cavity prior to the evacuation depends on the shift of the laser frequency in a succinct manner.

### A. Assessment of gas density by drift-free Fabry-Perot-based optical refractometry with no relocking

*1. Simplified expression for gas density in terms of the frequency shift of a laser locked to a mode of a drift-free optical cavity as the cavity is evacuated*

As is shown in the supplementary material (part 4), inserting Eq. (13) into Eq. (16) or (17) yields, for the case with no relocking, an expression for the density of gas in the cavity prior to the evacuation, $\rho_{n,i}$, that is expressed in a series of terms containing successively higher order powers of $\Delta \nu_l / \nu_0$, *viz.* as

$$\rho_{n,i} = \rho_{n,f} + \frac{2}{3A_R}\tilde{\chi}_{\Delta\nu}\frac{\Delta\nu_l}{\nu_0}\left[1 + \tilde{B}_{\Delta\nu}\frac{\Delta\nu_l}{\nu_0} + \tilde{C}_{\Delta\nu}\left(\frac{\Delta\nu_l}{\nu_0}\right)^2 + ...\right], \quad (18)$$

where $\rho_{n,f}$, which represents the density of the gas after the evacuation, henceforth denoted the residual gas density, is given by $2(n_f-1)/(3A_R)$, $\tilde{\chi}_{\Delta\nu}$ is a dimensionless factor that is close to unity, while $\tilde{B}_{\Delta\nu}$ and $\tilde{C}_{\Delta\nu}$ are normalized coefficients. All of the latter three, which are explicitly given in Table 2, depend, to various degrees, on the cavity deformation parameter, $\varepsilon$, the dispersion of the mirrors and the gas, through the $\varsigma$ and $\eta$ entities, the residual gas refractivity, i.e. $n_f - 1$, as well as the $A_R$, $B_R$, and $C_R$ virial coefficients. It is here of importance to note that the $\tilde{B}_{\Delta\nu}$ and $\tilde{C}_{\Delta\nu}$ coefficients do not only comprise the effects of the two- and three-body interactions and the non-linearity of the Lorentz-Lorenz equation, they are also affected by cavity deformation, dispersion effects, the refractivity of the residual gas as well as the continuous change of the frequency of the cavity mode as the cavity is evacuated. This implies that the leading terms of the $\tilde{B}_{\Delta\nu}$ and $\tilde{C}_{\Delta\nu}$ coefficients are dissimilar to those of the $\tilde{B}_\rho$ and $\tilde{C}_\rho$ coefficients.

Table 2. Explicit expressions for the various coefficients used for assessment of gas density from the shift of the frequency of a laser locked to a mode of a drift-free optical cavity as the cavity is evacuated with no relocking by use of Eq. (18) in terms of the cavity deformation parameter, $\varepsilon$, the dispersion parameters, $\varsigma$ and $\eta$, the virial coefficients, as well as the refractivity of the residual gas, $n_f - 1$.*

| Coefficient | Relation to the cavity deformation parameter, $\varepsilon$, the dispersion parameters, $\varsigma$ and $\eta$, the residual gas refractivity, $n_f - 1$, as well as the $A_R$, $B_R$, and $C_R$ virial coefficients. |
|---|---|
| $\tilde{\chi}_{\Delta\nu}$ | $\Omega(\varepsilon,\varsigma,\eta) + \frac{5}{3}(n_f-1)\left(1 - \frac{4}{5}\frac{B_R}{A_R^2}\right)$ |
| $\tilde{B}_{\Delta\nu}$ | $\frac{5}{6}\left[\Omega(\varepsilon,\varsigma,\eta)\left(1 - \frac{4}{5}\frac{B_R}{A_R^2}\right) \right.$ $\left. -\frac{8}{5}(n_f-1)\left(\frac{1}{24} + \frac{5}{6}\frac{B_R}{A_R^2} - \frac{4}{3}\frac{B_R^2}{A_R^4} + \frac{C_R}{A_R^3}\right)\right]$ |
| $\tilde{C}_{\Delta\nu}$ | $\frac{4}{9}\left(1 - \frac{5}{2}\frac{B_R}{A_R^2} + 2\frac{B_R^2}{A_R^4} - \frac{C_R}{A_R^3}\right)$ |

* Since $\varepsilon$, $\varsigma$, $\eta$, as well as $n_f - 1$ can be considered $\ll 1$, it can be estimated, by use of the information given in the footnote of Table 1, that the $\tilde{B}_{\Delta\nu}$ and $\tilde{C}_{\Delta\nu}$ coefficients for $N_2$ at 632 nm take values of 0.81 and -0.12, respectively.

Equation (18) thus constitutes the basis for assessments of gas density from a measurement of the shift of the cavity mode addressed by the laser light that follows an evacuation of the cavity in DF-FPC-OR systems in the absence of relocking.[41]

*2. Influence of the higher order terms on the assessment of the gas density in terms of the frequency shift of a laser locked to a mode of a drift-free optical cavity as the cavity is evacuated*

It was estimated above that, for a well-designed closed cavity, $\varepsilon$ can take values around $10^{-4}$ (although it can be an order of magnitude larger, i.e. up to $10^{-3}$, for an open cavity), $\varsigma$ can take a variety of values, ranging all from $10^{-6}$ to $10^{-3}$ (depending on the dispersion of the mirrors), while $\eta$ is typically around $10^{-6}$. It was moreover concluded that the cavity can presumably be evacuated so that the refractivity of the residual gas is not larger than few times $10^{-10}$. This implies that when accuracies down to $10^{-5}$ are needed, $\tilde{\chi}_{\Delta\nu}$ is basically given by $1 - \varepsilon + 2\varsigma$, while is should be given by the fully expression of $\Omega(\varepsilon,\varsigma,\eta)$, and, if the residual gas density is not extremely low, possibly also the $(n_f - 1)$ term, when accuracies down to $10^{-8}$ are required.

Moreover, since typical values for $\tilde{B}_{\Delta\nu}$ and $\tilde{C}_{\Delta\nu}$ are in the order of unity or below (for $N_2$ at 632 nm, 0.81 and -0.12, respectively), and since the relative shift of a mode of a cavity that is filled with gas under STP conditions that follows an evacuation is around a few times $10^{-4}$ ($3 \times 10^{-4}$ for $N_2^{STP}$ at 1.5 μm), Eq. (18) reveals that the first of the higher order terms within the set of brackets needs to be taken into account whenever assessments of gas density close to $\rho_n^{STP}$ conditions are to be performed with a (relative) accuracy that is better than around $10^{-4}$ ($2 \times 10^{-4}$ for $N_2^{STP}$). The second non-linear term needs to be included if the density is to be determined with a (relative) accuracy that is several orders of magnitude better than this, i.e. in the $10^{-8}$ range ($1 \times 10^{-8}$ for $N_2^{STP}$). Hence it can be concluded that the last term needs to be taken into account under STP conditions solely when $A_R$ and $\tilde{B}_{\Delta\nu}$ are known with (exceptionally) high accuracy or when accurate calibrations are to be made.

As before, if gas densities below those at STP are to be assessed, $\Delta\nu_l$ becomes smaller, whereby the higher order terms play a successively smaller role, while, if gas densities above $\rho_n^{STP}$ are to be assessed, the higher order terms play a progressively larger role.

**B. Assessment of gas density by drift-free Fabry-Perot-based optical refractometry in the presence of relocking**

*1. Simplified expression for gas density in terms of the shifts of frequency and mode number of a laser locked to a mode of a drift-free optical cavity as the cavity is evacuated*

As is shown in the supplementary material (part 4), inserting Eq. (14) into Eq. (16) or (17) yields, for the case with relocking, an expression for the density of gas in the cavity prior to the evacuation, Eq. (S55), that reads

$$\rho_{n,i} = \rho_{n,f} + \frac{2}{3A_R}\tilde{\chi}_{\Delta q}\left\{\frac{\Delta q}{q_0}\left[1 + \tilde{B}_-^{(\Delta q)^2}\frac{\Delta q}{q_0} + \tilde{C}_-^{(\Delta q)^3}\left(\frac{\Delta q}{q_0}\right)^2\right] \right.$$
$$+ \frac{\Delta\nu_l}{\nu_0}\left[\tilde{A}_{\Delta\nu}^- + \tilde{B}_{\Delta\nu}^{\Delta q}\frac{\Delta q}{q_0} + \tilde{C}_{\Delta\nu}^{(\Delta q)^2}\left(\frac{\Delta q}{q_0}\right)^2\right]$$
$$\left. + \left(\frac{\Delta\nu_l}{\nu_0}\right)^2\left[\tilde{B}_{(\Delta\nu)^2}^- + \tilde{C}_{(\Delta\nu)^2}^{\Delta q}\frac{\Delta q}{q_0}\right]\right\}, \quad (19)$$

where the $\tilde{\chi}_{\Delta q}$ factor is a dimensionless factor that is close to unity while the $\tilde{B}_-^{(\Delta q)^2}$, $\tilde{C}_-^{(\Delta q)^3}$, $\tilde{A}_{\Delta\nu}^-$, $\tilde{B}_{\Delta\nu}^{\Delta q}$, $\tilde{C}_{\Delta\nu}^{(\Delta q)^2}$, $\tilde{B}_{(\Delta\nu)^2}^-$, and $\tilde{C}_{(\Delta\nu)^2}^{\Delta q}$ are various normalized coefficients. Table 3 provides explicit expressions for all these entities in terms of the $A_R$, $B_R$, and $C_R$ virial coefficients, the cavity deformation parameter, $\varepsilon$, the dispersion parameters, $\varsigma$ and $\eta$, and the residual gas refractivity, i.e. $n_f - 1$. It is worth to note that although the $\tilde{\chi}_{\Delta q}$ factor depends on $\varepsilon$, it is not affected by dispersion, i.e. $\varsigma$ and $\eta$.

*2. Influence of the higher order terms on the assessment of the gas density in terms of the shifts of frequency of a laser locked to a mode and the change in mode number of a drift-free optical cavity as the cavity is evacuated*

As was alluded to above, when relocking takes place, it is customary to relock the laser to a mode so that $\Delta\nu_l/\nu_0$ is below $\nu_{FSR}^0/\nu_0$ whereby $\Delta q/q_0$ is in the order of $n_i - 1$. This implies that under STP conditions $\Delta q/q_0$ normally takes a value of a few times $10^{-4}$ (around $3 \times 10^{-4}$ for $N_2^{STP}$ at 1.5 μm) while $\Delta\nu_l/\nu_0$ is significantly smaller; it can take any value from zero to $\nu_{FSR}^0/\nu_0$ ($4 \times 10^{-6}$ for a cavity with a length of 20 cm and light with a wavelength of 1.5 μm).

Since the various coefficients of the higher order terms in Eq. (19) typically range from $10^{-1}$ to $10^{1}$ (for $N_2$ at 632 nm as given by the footnote of Table 3), it is possible to conclude that in most cases, and in particular under STP conditions, the largest non-linearity originates from the first non-unity term within the first set of brackets, i.e. the $\tilde{B}_{-}^{(\Delta q)^2}(\Delta q / q_0)$ term, which adds a contribution that is in the order of a few times $10^{-5}$ ($6 \times 10^{-5}$ for $N_2^{STP}$ at 632 nm). The next order term within the same set of brackets, i.e. the $\tilde{C}_{-}^{(\Delta q)^3}(\Delta q / q_0)^2$ term, contributes significantly less; again for the case with $N_2^{STP}$ at 632 nm, almost three orders of magnitude less, ca. $7 \times 10^{-8}$. On the other hand, the contribution of the first order term within the second set of brackets, i.e. the $\tilde{B}_{\Delta \nu}^{\Delta q}(\Delta q / q_0)$ term, depends on the actual size of the $\Delta \nu_l / \nu_0$ term. For the case when $\Delta \nu_l / \nu_0$ is close to $\nu_{FSR}^0 / \nu_0$, it contributes to the measured density with *ca.* $10^{-6}$. Under the same conditions, the $\tilde{B}_{(\Delta \nu)^2}^{-}$ term within the third set of brackets, which corresponds to the most prominent non-linear contribution in the case with no relocking, amounts in this case to only *ca.* $2 \times 10^{-9}$. All this shows that the non-linear effects are dissimilar in the cases with and without relocking, and more pronounced when DF-(D)FPC-OR is made without relocking than with.

Table 3. Explicit expressions for the various factors and coefficients used for assessment of gas density in terms of the shifts of the frequency of a laser locked to a mode and the change in mode number addressed by the laser light of a drift-free optical cavity as the cavity is evacuated in the presence of relocking by use of Eq. (19).*

| Coefficient | Relation to the cavity deformation parameter, $\varepsilon$, the dispersion factors, $\varsigma$ and $\eta$, the residual gas refractivity, $n_f - 1$, as well as the $A_R$, $B_R$, and $C_R$ virial coefficients. |
|---|---|
| $\tilde{\chi}_{\Delta q}$ | $(1 - \varepsilon + \varepsilon^2) + (n_f - 1)\frac{2}{3}\left[\left(1 - \frac{5}{2}\varepsilon + \frac{5}{2}\varepsilon^2\right) - 2\frac{B_R}{A_R^2}(1 - \varepsilon + \varepsilon^2)\right]$ |
| $\tilde{A}_{\Delta \nu}^{-}$ | $(1 + 2\varsigma + \eta) + \Omega(\varepsilon, \varsigma, \eta)(n_f - 1)$ |
| $\tilde{B}_{-}^{(\Delta q)^2}$ | $-\frac{1}{6}\left[(1 + 5\varepsilon) + 4(1 - \varepsilon)\frac{B_R}{A_R^2}\right]$ $-(n_f - 1)\frac{8}{9}\left[1 - \frac{1}{16}\varepsilon + \frac{1}{2}(1 - \frac{5}{2}\varepsilon)\frac{B_R}{A_R^2}\right.$ $\left.-2(1 - \varepsilon)\frac{B_R^2}{A_R^4} + \frac{3}{2}(1 - \varepsilon)\frac{C_R}{A_R^3}\right]$ |
| $\tilde{C}_{-}^{(\Delta q)^3}$ | $-\frac{2}{9}\left(1 - \frac{B_R}{A_R^2} - 4\frac{B_R^2}{A_R^4} + 2\frac{C_R}{A_R^3}\right)$ |
| $\tilde{B}_{\Delta \nu}^{\Delta q}$ | $\frac{2}{3}\left[\left(1 - \frac{5}{2}\varepsilon + 2\varsigma + \eta\right) - 2\Omega(\varepsilon, \varsigma, \eta)\frac{B_R}{A_R^2}\right]$ $-(n_f - 1)\frac{10}{9}\left[\left(1 + \frac{4}{5}\varepsilon + 2\varsigma + \eta\right) + 2\Omega(\varepsilon, \varsigma, \eta)\frac{B_R}{A_R^2}\right.$ $\left.-\frac{16}{5}\Omega(\varepsilon, \varsigma, \eta)\frac{B_R^2}{A_R^4} + \frac{12}{5}\Omega(\varepsilon, \varsigma, \eta)\frac{C_R}{A_R^3}\right]$ |
| $\tilde{C}_{\Delta \nu}^{(\Delta q)^2}$ | $-\left(1 + \frac{2}{3}\frac{B_R}{A_R^2} - \frac{8}{3}\frac{B_R^2}{A_R^4} + \frac{4}{3}\frac{C_R}{A_R^3}\right)$ |
| $\tilde{B}_{(\Delta \nu)^2}^{-}$ | $\frac{5}{6}\left[(1 - \varepsilon + 4\varsigma + 2\eta) - \frac{4}{5}(1 - \varepsilon + 4\varsigma + 2\eta)\frac{B_R}{A_R^2}\right]$ $+(n_f - 1)\frac{7}{9}\left[\left(1 + \frac{1}{14}\varepsilon + 4\varsigma + \eta\right) - \frac{16}{7}(1 - \frac{5}{8}\varepsilon + 4\varsigma + 2\eta)\frac{B_R}{A_R^2}\right.$ $\left.+\frac{16}{7}(1 - \varepsilon + 4\varsigma + 2\eta)\frac{B_R^2}{A_R^4} - \frac{12}{7}(1 - \varepsilon + 4\varsigma + 2\eta)\frac{C_R}{A_R^3}\right]$ |
| $\tilde{C}_{(\Delta \nu)^2}^{\Delta q}$ | $-\frac{1}{3}\left(1 + 6\frac{B_R}{A_R^2} - 8\frac{B_R^2}{A_R^4} + 4\frac{C_R}{A_R^3}\right)$ |

\* By use of the information given in the footnote of Table 1, the factors and coefficients given above for $N_2^{STP}$ at 632 nm can be estimated to take the following values: $\tilde{\chi}_{\Delta q} = 1$; $\tilde{A}_{\Delta \nu}^{-} = 1$; $\tilde{B}_{-}^{(\Delta q)^2} = -0.19$, $\tilde{C}_{-}^{(\Delta q)^3} = -0.75$, $\tilde{B}_{\Delta \nu}^{\Delta q} = 0.62$, $\tilde{C}_{\Delta \nu}^{(\Delta q)^2} = -2.6$, $\tilde{B}_{(\Delta \nu)^2}^{-} = 0.81$, and $\tilde{C}_{(\Delta \nu)^2}^{\Delta q} = -10.5$.

As before, if gas densities below those at STP are to be assessed, $\Delta q$ becomes smaller whereby the higher order terms in the first set of brackets, i.e. $\tilde{B}_{-}^{(\Delta q)^2}(\Delta q / q_0)$ and $\tilde{C}_{-}^{(\Delta q)^3}(\Delta q / q_0)^2$, play successively smaller roles, while, if gas densities above $\rho_n^{STP}$ are to be assessed, they play a correspondingly larger role. The higher order terms in the second brackets, on the other hand, do not primarily depend on the total gas density probed, but rather how much the frequency of the laser needs to shift to find a new mode to which it can lock. Hence, they can shift from zero to the values they take when $\Delta \nu_l / \nu_0$ is around $\nu_{FSR}^0 / \nu_0$, as given above.

## VI. SUMMARY AND CONCLUSIONS

Optical Refractometry (OR) is a technique for assessment of refractivity that has demonstrated an exceptionally high precision. Although it can be performed by various means, it has been found that the use of resonant Fabry-Perot (FP) cavities have significant advantages over other means, e.g. use of Michelson interferometers.[14] Since density is predominantly linearly dependent of refractivity, and this dependence does not have any temperature dependence, FP Cavity-based OR (FPC-OR) has a great potential for both high precision and accurate assessment of gas (number or molecular) density under a variety of conditions.

However, as is discussed in some detail in one of our two accompanying papers,[27] although both refractivity and gas density can be assessed with exceptional high precision, they cannot, in general, be determined with as high accuracy. One reason for this is the lack of standard reference refractivity and gas density sources. For a system that has not been calibrated to any standard, the accuracy is mainly limited by the relative accuracy by which the polarizability, $A_R$, is known, which for nitrogen presently is in the $10^{-5}$ - $10^{-4}$ range. If such a system could be characterized by one type of gas whose refractivity or gas density is well-known under at least one specific set of conditions, (i.e. known with a low uncertainty), which is an issue that is to be discussed in a future work, the system would demonstrate a significantly better accuracy. In this case, the main reason for uncertainty in an assessment is changes of the length of the cavity, primarily originating from various types of drifts of the length of the cavity spacer, including creeping and relaxations.

A possible means to minimize these types of effects is to use two FP cavities, bored in the same cavity spacer block, here referred to as a Dual Fabry-Perot Cavity (DFPC) and denoted DFPC-OR. By locking two lasers, each to a longitudinal mode of its own cavity, it is assumed that the differential drift in length between the two cavities will be reduced, thus minimizing the effects of thermal drifts and relaxations on the assessment. This has also the advantage that the two laser beams can be mixed on a photo detector to produce a beat signal, whereby the shift in frequency of one of the beam can be conveniently measured as the shift in beat frequency. However, despite the use of DFPC it has been found that the measurements are still often limited by residual differential drifts of the lengths of the two cavities. Much would therefore be gained if the technique could be realized under drift-free conditions, here denoted DF-FPC-OR (or DF-DFPC-OR), since, under such conditions, the technique would be virtually unaffected by the limitations imposed by the cavity spacer material.

Although it recently has been reported on a DFPC-OR system with a minimum of drifts and relaxations, obtained by long-term baking of a dual-cavity spacer placed in a temperature controlled compartment,[19] we have chosen to alleviate these types of problems by an alternative means; *viz.* by developing novel methodologies for DFPC-OR that are based on *fast switching* of gas volumes, denoted FS-FPC-OR (or FS-DFPC-OR). By this, it is prophesized that (D)FPC-OR can be utilized for high precision and accurate assessments of gas density under drift-free cavity conditions under a wide range of conditions. Various means and methodologies to realize FS-DFPC-OR are presented and further discussed in one of our accompanying papers.[26]

However, it has been concluded that the FPC-OR technique is, in reality, affected by a variety of phenomena and non-linear responses (some of which have technical origin) that need to be taken into account properly if assessments of refractivity and gas density are to be made with high accuracy. To be able to use a DF-(D)FCP-OR instrumentation for such assessments, it is in general not sufficient to use the simplest (first order) and most common expressions for how the frequency of laser light that is locked to a mode of the cavity is related to the refractivity and the density of the gas, and, in particular, how this is changed, i.e. $\Delta \nu_l$, as the cavity is evacuated, that are valid for an ideal gas assessed by the use of (an) ideal cavity (ies), e.g. by either of the Eqs. (3) or (5) and the Eqs. (8) or (9) given in this work. Instead, more general (but still explicit) expressions are needed.

This paper, which thus constitutes the first in a series of three papers dealing with DF-(D)FPC-OR in general, and FS-(D)FPC-OR in particular, has presented such expressions, based on relations of how gas density is related to the refractivity of the gas, $n-1$, and how the latter depends on $\Delta \nu_l$, that properly acknowledges the influence of all conceivable phenomena and higher order (i.e. non-linear) contributions to these expressions.

For the assessment of refractivity, processes that have been taken into account comprise the continuous change of the frequency of the cavity mode as the cavity is evacuated, the instantaneous deformation of the cavity, the penetration of light into the coating of the mirrors, including its dispersion, the dispersion of the gas, and the influence of residual gas density in the cavity following an evacuation. For the assessment of gas density, which is built upon the assessment of refractivity (and thereby include the aforementioned processes), additionally also the Lorentz-Lorenz equation and description of the state of the gas in terms of a series of higher order terms, including virial coefficients, have been taken into account. The influences of all these effects on FPC-OR have been individually covered and their combined effects of DF-FPC-OR are described in a coherent manner. Simple and explicit expressions have been derived for both the cases with and without relocking of the laser to other cavity modes that take all these phenomena into account. Comparisons with the first order expressions, i.e. those derived for a thin ideal gas in an ideal optical cavity, have been made.

It has been concluded, and as is further discussed in one of our accompanying paper,[27] that the processes and non-linearities that play the largest roles for the accuracy of an OR-assessment of either refractivity or gas density by an instrumentation that has been calibrated by one type of gas whose refractivity or gas density is well-known under at least one specific set of conditions, are *i)* the fact that the frequency of the laser locked to the cavity shifts continuously as the gas density in being changed, *ii)*, the deformation of the cavity due to the pressure of the gas, *iii)* the 2$^{nd}$ virial coefficient in the extended Lorentz-Lorenz equation, $B_R$, *iv)* the dispersion of the mirrors and the gas, and *v)* the refractivity of the residual gas. Unless all these are included in the expressions for gas density and refractivity, the accuracy of the assessments can be compromised.

Based on all this, it has been concluded that neglect of the most prominent non-linear effects can, for a well-designed closed cavity system, affect the accuracy of the technique under standard pressure and temperature (STP) conditions with up to around $10^{-4}$ when well-designed closed cavities are being used (and up to an order of magnitude more for poorly designed cavities or open ones). Hence, for the highest accuracies, such non-linearities need therefore to be accounted for, both in the characterization and the measurement procedure.

In addition to the expressions given and the aforementioned conclusions, the analysis of the phenomena that potentially can affect DF-(D)FPC-OR performed has made it possible to draw a few other important conclusion and provide a number of guidelines for how to construct DF-(D)FPC-OR instrumentation with the highest accuracy, of which the most important ones are:

1) the influence of the cavity deformation is larger for an open cavity than a closed one;
2) the non-linearities from dispersion are smaller for the cases when relocking takes place than when it is not used; and
3) there is an advantage if mirror with a coating for which the outermost layer have the highest index of refraction can be used.

The reason for why the cavity deformation is larger for an open cavity than a closed one is that in the former case the gas affects the spacer material with a significantly larger force than in the latter; an open cavity is basically affected by a force that is given by the product of the gas pressure and the cross section of the entire spacer, while, for a closed cavity, the force is solely given by the product of the gas pressure and the diameter of the area of the mirrors that are exposed to a gas (which often is equal to the area of the cavity).

The reason for the fact that non-linearities from dispersion are smaller for the cases when relocking takes place than when it is not used is that such non-linearities depend on the shift of the frequency of the light field. In the case with relocking, the change in frequency of the light is smaller than that when no relocking is used.

As is alluded to in the supplementary material (part 2), the reason for the advantage of mirrors with coatings for which the outermost layer have the highest index of refraction is that for such coatings the penetration depth is proportional to the index of refraction inside the cavity.[42] This implies, in contrast to what previously has been mentioned,[19] that the light in the cavity experiences the index of refraction of the gas over the entire effective length of the cavity. Hence, the finite penetration depth of the mirrors does not have any noticeable effect on the accuracy of the technique. This is in contrast to the case for mirrors for which the outermost layer has a lower index of refraction than the next layer.[42] The coating configuration considered, with the outermost layer having the highest index of refraction, is therefore the one preferred for FPB-OR if mirror dispersion effects are limiting the instrumentation.

In conclusion, the expressions derived in this work can serve as a basis for assessments of refractivity [the Eqs. (13) and (14)] as well as gas density [the Eqs. (18) and (19)], and changes in such, by the DF-FPC-OR (or DF-DFPC-OR) technique in such a way that the assessments can benefit maximally from the extraordinary power of FPC-OR. They can also serve as basis for the development of powerful methodologies for assessments of refractivity and gas density, and changes in such, by FS-DFPC-OR. Such methodologies are suggested in one of our accompanying papers.[26] Finally, the expected precision, accuracy, and temperature dependence that can be obtained from DF-FPC-OR in general, and these methodologies in particular, are specifically addressed in a separate paper.[27] By all this, we consider that this work, which serves as the basis for DF-FPC-OR and DF-DFPC-OR, has laid the foundation for future constructions of DF-FPC-OR and DF-DFPC-OR instrumentations that can be used for precise and accurate assessment of gas density with high accuracy under a variety of conditions, including the important non-temperature stabilized situations. It is also assumed that it will have a significant impact on the development of gas density standards.

## SUPPLEMENTARY MATERIAL

See the supplementary material for a derivation of simple expressions for the refractivity and density of a thin ideal gas in terms of the frequency shift of a mode of an ideal optical cavity and that of a laser field locked to such a mode as the cavity is evacuated (part 1); an assessment of refractivity by measurement of the shift of the frequency of a laser field locked to a mode of a drift-free optical cavity as the cavity is evacuated in the presence of an instantaneous cavity deformation, light penetration into the mirror coatings, and gas dispersion (part 2); a derivation of an expression for the density of a real gas in terms of refractivity (part 3); and an assessment of the density of molecules by measurement of the frequency shift of a laser field locked to a mode of a drift-free optical cavity as the cavity is evacuated (part 4).

## Acknowledgment

This research was supported by the EMPIR initiative, which is co-founded by the European Union's Horizon 2020 research and innovation program and the EMPIR Participating States; the Swedish Research Council (VR), project numbers 621-2011-4216 and 621-2015-04374; Umeå University program VFS (Verifiering för samverkan), 2016:01; and Vinnova Metrology Programme, project number 2015-0647 and 2014-06095.

the cavity modes and that of the laser frequencies, this will not, in any way, affect the results presented or the conclusions drawn.

[32] To conform with the SI system, even if $\alpha$ and $\chi$ are given in units of cm$^3$, it is customary to give $A$ in units of m$^3$, since the density then naturally will be given in number of molecules per cubic meter. In this case, $A$ is given by $(4\pi/3)10^{-6}(\alpha+\chi)$. An alternative is to express the density as a molar density, i.e. in units of moles per unit volume (which is the inverse of the molar volume). In such a case, Eq. (7) needs to be adjusted by Avogadro's number, $N_A$. For the case when $A$ is to be expressed in units of m$^3$/mol, it should thereby be assessed as $(4\pi/3)10^{-6}N_A(\alpha+\chi)$. Irrespective of whether the density is given in units of number of molecules or moles per cubic meter, it will in this work be denoted $\rho_n$. Its dimensions are in all cases given by the inverse of those of $A$.

[33] K. Piszczatowski, M. Puchalski, J. Komasa, B. Jeziorski and K. Szalewicz, Phys. Rev. Lett. **114** (17), 173004 (2015).

[34] P. J. Mohr, B. N. Taylor and D. B. Newell, J. Phys. Chem. Ref. Data **41** (4), 1527-1605 (2012).

[35] We have here neglected the influence of the phase development a focused laser beam experiences along its propagation direction due to its smallness and the fact that it is not affected, to first order, by the evacuation process.[39,43]

[36] The reason why $\varepsilon$ is significantly larger for an open cavity placed in a compartment in which gas is introduced than for a closed cavity in which the gas solely is introduced in the cavity is that in the former case the gas affects the spacer material with a significantly larger force than in the latter. An open cavity is basically affected by a force that is given by the product of the gas pressure and the cross section of the entire spacer, while, for a closed cavity, the force is solely given by the product of the gas pressure and the diameter of the area of the mirrors that are exposed to a gas (which often is equal to the area of the cavity). This implies that $\varepsilon$ can be significantly smaller for a closed cavity than for an open (by up to an order of magnitude).

[37] Although the mirror dispersion parameter, $\zeta$, can be of similar magnitude as the cavity deformation parameter, $\varepsilon$, and the latter has been expanded to second order, we have chosen to not do the same for $\zeta$. The reason is that although $\zeta$ in principle can take such values, it is presumably seldom mirrors with such coatings will be used (since it is nowadays practically fairly simple to obtain mirrors with low amount of dispersion).

[38] A. Hourri, J. M. Starnaud and T. K. Bose, J. Chem. Phys. **106** (5), 1780-1785 (1997).

[39] E. R. Peck and B. N. Khanna, J.Opt. Soc. Am. **56** (8), 1059-1063 (1966).

[40] Although the experimental work performed in our accompanying paper[26] was performed at 1.5 µm, since very few virial coefficients have so far been assessed in this wavelength region, we will, in this work, whenever various entities are estimated, unless otherwise stated, use the virial coefficients of N$_2$ that are assessed at 632 nm, given in the footer of Table 1. This implies that some of the estimates given here might not be fully appropriate for all possible wavelengths at which FPC-OR can be performed. However, since they are assumed to reflect typical values, they are in this work considered to provide relevant order-of-magnitude estimates of the relative size of various terms in the expressions derived.

[41] In all expressions given as series expansions, e.g. the Eqs (15) - (18), as many terms that are needed to obtain the required accuracy should be included.

[42] L. R. Brovelli and U. Keller, Opt. Commun. **116** (4-6), 343-350 (1995).

# Supplementary material to "Drift-free Fabry-Perot-cavity-based optical refractometry – accurate expressions for assessments of gas refractivity and density"

Ove Axner,[1,a)] Isak Silander,[1] Thomas Hausmaninger,[1] and Martin Zelan[2]

[1]*Department of Physics, Umeå University, SE-901 87 Umeå, Sweden*

[2]*Measurement Science and Technology, RISE Research Institutes of Sweden, SE-501 15 Borås, Sweden*

**Part 1 – Derivation of simple expressions for the refractivity and density of a thin ideal gas in terms of the frequency shift of a mode of an ideal optical cavity and that of a laser field locked to such a mode as the cavity is evacuated**

*Part 1.1. A simple expression for the frequency shift of a mode of an ideal optical cavity as the cavity is evacuated*

In the simplest of pictures it is often assumed that the frequency, $\nu_q$, of the $q$:th longitudinal mode of a Fabry-Perot cavity containing a gas with an index of refraction of $n$ can be expressed as

$$\nu_q(n) = \frac{qc}{2nL_0},  \quad (S1)$$

where $c$ is the speed of light in vacuum and $L_0$ the length of the empty cavity. This implies that it is often concluded that, as the cavity is evacuated, the frequency of this mode is shifted an amount $\Delta\nu_{cm}$ that is given by[1]

$$\begin{aligned}\Delta\nu_{cm} &= \nu_q(n=1) - \nu_q(n) = \frac{qc}{2L_0} - \frac{qc}{2nL_0} = \nu_0\left(1 - \frac{1}{n}\right) \\ &= \nu_0\left[1 - \frac{1}{1+(n-1)}\right] \\ &= \nu_0\left\{1 - \left[1-(n-1)+(n-1)^2-(n-1)^3+O[(n-1)^4]\right]\right\} \\ &= \nu_0\left\{(n-1)-(n-1)^2+(n-1)^3+O[(n-1)^4]\right\},\end{aligned} \quad (S2)$$

where $\nu_0$ is the frequency of the cavity mode of an empty cavity and where we have series expanded the term comprising the refractivity. However, it is important to note that Eq. (S2) assumes that all entities in Eq. (S1), except the index of refraction, are constants during the evacuation process. This implies implicitly that both the gas and the cavity are treated as ideal.

This shows that under the condition that the gas additionally is considered to be thin, i.e. when $n-1 \ll 1$, it is possible to neglect all higher order terms in Eq. (S2), which leads to Eq. (2) that states that

$$\Delta\nu_{cm} = \nu_0(n-1). \quad (S3)$$

For the case when the frequency of a laser is locked to the same cavity mode during the entire gas evacuation process, the shift of the frequency of the laser light, $\Delta\nu_l$, is the same as that of the cavity mode, $\Delta\nu_{cm}$. This implies that the refractivity of the gas often is considered to be given by

$$n-1 = \frac{\Delta\nu_l}{\nu_0}. \quad (S4)$$

When the laser field is relocked to a different mode during the evacuation, the shift of the frequency of the laser is given by

$$\Delta\nu_l = \Delta\nu_{cm} - \Delta q \cdot \nu_{FSR}^0, \quad (S5)$$

where $\Delta q$ represents the number of modes by which the laser field is shifted during the relocking process, while $\nu_{FSR}^0$ is the free spectral range of the empty cavity, given by $c/(2L_0)$. This implies the refractivity often is considered to be given by

$$n-1 = \frac{\Delta\nu_l}{\nu_0} + \frac{\Delta q}{q_0}, \quad (S6)$$

where $q_0$ is the number of the mode addressed, alternatively given by $\nu_0/\nu_{FSR}^0$.

This shows that the Eqs (S4) and (S6) in reality are the first order terms of more general expressions and that they are, in principle, valid solely when both the gas and the cavity are ideal (and thus that the gas is infinity thin). They should therefore only be used when low accurate assessments are to be performed or possibly when small gas densities are considered (when the higher order terms are sufficiently small so they can be neglected with respect to the required accuracy).

*Part 1.2. A simple expression for the density of a thin ideal gas in terms of refractivity*

Under the condition that the Lorentz-Lorenz equation is valid and that the gas can be assumed to be ideal, it is possible to relate the index of refraction, $n$, to the density of molecules, $\rho_n$, according to (in CGS units)

$$\frac{n^2-1}{n^2+2} = \frac{4\pi}{3}(\alpha + \chi)\rho_n, \quad (S7)$$

where $\alpha$ and $\chi$ are the (dipolar) polarizability and the diamagnetic susceptibility, respectively.[2]

Again for the case with a thin gas, for which the refractivity is small (i.e. when $n-1 \ll 1$), it is possible to expand the left hand side of Eq. (S7) around $n-1$, which gives

$$\frac{n^2-1}{n^2+2} \approx \frac{2}{3}(n-1)\left[1 - \frac{1}{6}(n-1) - \frac{2}{9}(n-1)^2 + ...\right]. \quad (S8)$$

This shows that for a thin ideal gas, it is often possible to assume that the density is given by

$$\rho_n = \frac{2}{3A}(n-1), \quad (S9)$$

where $A$, which often is termed the (number or molar) polarizability, is given by

$$A = \frac{4\pi}{3}(\alpha + \chi). \quad (S10)$$

However, as can be seen from Eq. (S8), Eq. (S9) is simply the first order term in a series expansion of the Lorentz-Lorenz equation. The neglect of the higher order terms in Eq. (S9) and the use of Eq. (S10), thereby limits the relative accuracy of Eq. (S10) for assessment of gas density to at least 1/6 of the refractivity of the gas being addressed, which under STP conditions (for which $n$ is typically $3 \times 10^{-4}$, see below) corresponds to $5 \times 10^{-5}$.

*Part 1.3. A simple expression for the density of a thin ideal gas in terms of the frequency shift of a mode of an ideal optical cavity and a laser locked to such a mode as the cavity is evacuated*

By use of the Eqs. (S3) and (S9) it is often assumed that $\rho_n$ can be assessed as

$$\rho_n = \frac{2}{3A}\frac{\Delta\nu_{cm}}{\nu_0}. \tag{S11}$$

This implies that for the case when the frequency of a laser is locked to the same cavity mode during the entire gas evacuation process, the gas density is often considered to be given by

$$\rho_n = \frac{2}{3A}\frac{\Delta\nu_l}{\nu_0}. \tag{S12}$$

For the case when the laser field is relocked to a different mode during the evacuation, the gas density is often considered to be given alternatively by

$$\rho_n = \frac{2}{3A}\left(\frac{\Delta\nu_l}{\nu_0} + \Delta q \cdot \frac{\nu_{FSR}^0}{\nu_0}\right) = \frac{2}{3A}\left(\frac{\Delta\nu_l}{\nu_0} + \frac{\Delta q}{q_0}\right). \tag{S13}$$

However, as is illustrated by the derivation of Eq. (S11), on which the latter two expression are based, these are just the first order approximations of more general expressions for assessments of gas density by FPC-OR. Despite this, they appear in the literature. This shows that it is justified to use them solely when low accurate assessments are to be performed or when small gas densities are considered (when the higher order terms are sufficiently small so they can be neglected with respect to the required accuracy).

**Part 2 – Assessment of refractivity by measurement of the shift of the frequency of a laser field locked to a mode of a drift-free optical cavity as the cavity is evacuated in the presence of an instantaneous cavity deformation, light penetration into the mirror coatings, and gas dispersion**

As was alluded to in section 3, to provide an accurate description of refractometry, it is necessary to include all relevant phenomena that can affect an FPC-OR assessment, even when it is performed under drift-free conditions.

There are a variety of processes that causes the physical dimensions of the cavity to change. In addition to the common ones, thermal drift and instantaneous cavity deformation due to the presence of gas, the length of a cavity mode can be altered also by other, primarily slow processes, e.g. drifts due to cavity mode creeping and relaxations in the material.[3-5] However, since DF-FCP-OR methodologies (including FS-DFPC-OR realizations presented in one of our accompanying works[6]) deal with measurements performed under drift-free conditions, they are not affected by slow processes. We will therefore, in this work, discard all such processes and solely consider instantaneous deformation processes.

This implies that the expression for the frequency, $\nu_q$, of the $q$:th longitudinal mode of a Fabry-Perot cavity containing a gas with an index of refraction of $n$, which for the simplest of cases was given by Eq. (S1), should be written as

$$\nu_q(n) = \frac{qc}{2\left[n(\nu_q)L(p) + 2L_{pd}(\nu_q)\right]}, \tag{S14}$$

where $L(p)$ is the physical length of the cavity (at a given pressure, $p$), defined as the distance between the surfaces of the mirrors (hence, the length over which gas can be filled), while $L_{pd}(\nu_q)$ is the penetration depth of the light into a mirror.[7] The entity within the set of brackets will henceforth be referred to as the effective optical cavity length and be denoted $L_{eff}$. Note that we have here neglected the influence of the phase development a focused laser beam experiences along its propagation direction due to its smallness and the fact that it is not affected, to first order, by the evacuation process.[8,9]

*Part 2.1. Effect of instantaneous cavity deformation on the frequency of a longitudinal cavity mode*

There are several instantaneous processes that cause the physical dimensions of the cavity to change [i.e. by producing an elongation $\Delta L_E(p)$], primarily those that are caused by the forces the gas exerts on the cavity spacer and the mirrors. The two most prominent ones originate from *i)* an instantaneous change in length of the cavity (often referred to as strain) due to stress in the spacer material, here referred to as spacer deformation and denoted $\Delta L_{CS}(p)$, and *ii)* bending of the mirrors, which gives rise to a displacement of the center of the mirror, referred to as $\Delta L_m(p)$. This implies that we will envision the physical length of the cavity to be given by

$$L(p) = L_0 + \Delta L_E(p) = L_0 + \Delta L_{CS}(p) + \Delta L_m(p), \tag{S15}$$

where, as before, $L_0$ is the physical length of the empty cavity.

**Spacer deformation:** Regarding the strain in the spacer material, it can be concluded that for an open cavity, which is affected by the gas from all directions,[3] the instantaneous contraction (negative elongation) of the cavity spacer that results when the cavity is filled with gas to a pressure $p$ can be estimated as [10]

$$\Delta L_{CS}(p) = -\frac{L_0}{3E}p, \tag{S16}$$

where $E$ is the Young's modulus of the cavity spacer material. Since this elongation is directly proportional to $p$, it represents an elastic response.

For a closed cavity, on the other hand, in which the gas primarily exerts a pressure on the mirrors from one direction (from inside the cavity, the pressure from outside is constant, given by the surrounding pressure),[5] the instantaneous elongation of the cavity spacer can be estimated from a similar but yet slightly different expression, given by

$$\Delta L_{CS}(p) = \frac{\beta L_0}{E}p, \tag{S17}$$

where $\beta$ is a dimensionless geometrical parameter that can be estimated to take a value between $A_{bore}/A_{CS}$ and $A_{bore}/A_{mirror}$, where $A_{bore}$, $A_{CS}$, and $A_{mirror}$ are the cross section areas of the bore, the cavity spacer, and the mirrors respectively.

This shows that irrespectively of whether the cavity is open or closed, the instantaneous change in length of the cavity spacer can be seen as a function of pressure, i.e. as $\Delta L_{CS}(p)$. Moreover, since the geometrical factor $\beta$ can be considered to be $< 0.1$ for a well-designed cavity, this also illustrates that the instantaneous elongation of the cavity spacer that takes place when the cavity is filled with gas can be considerably smaller for a closed cavity than for an open one. Since $E$ often is in the order of 90 GPa (for Zerodur) the relative length deformation of the spacer, $\Delta \bar{L}_{CS}(p)$, defined as $\Delta L_{CS}(p)/L_0$, will, under STP conditions, be around $4 \times 10^{-7}$ for an open cavity, while it will be significantly smaller for a well-designed closed cavity; around $4 \times 10^{-8}$ for the case when $\beta = 0.03$.

**Bending of mirrors:** The other main contribution to the instantaneous cavity elongation, $\Delta L_m(p)$, originates from bending of the mirrors. The distance the center of the mirror is shifted due to bending does not only depend on the properties of the mirror substrate, but also how the mirror is attached to the spacer.[S1] For mirrors that can be described as uniformly loaded circular plates with clamped edges, it is possible to estimate the shift of its center position by [11]

$$\Delta L_m(p) = \frac{A_{bore}^2}{\pi^2} \frac{3(1-\bar{v})}{32Gh^3} p ,\qquad (S18)$$

where $\bar{v}$, $G$, and $h$ are the Poisson's ratio, the shear modulus, and the thickness of the mirror substrate, respectively. For other situations, similar but slightly different types of expressions are valid.

This shows that also the bending of the mirrors can be seen as a function of pressure, i.e. as $\Delta L_m(p)$. As is shown in the footnote below, under STP conditions, $\Delta \bar{L}_m(p)$, defined as $\Delta L_m(p)/L_0$, can typically be in the order of $10^{-10}$.[S2] This implies that $\Delta L_m(p)$ is in most cases smaller than $\Delta L_{CS}(p)$, in particular for an open cavity. It is therefore possible to conclude that under most conditions, $\Delta L_E(p) \approx \Delta L_{CS}(p)$.

***Total relative instantaneous elongation of the cavity:*** The total relative instantaneous elongation of the cavity, $\Delta \bar{L}_E(p)$, defined as $\Delta L_E(p)/L_0$, which, in the general case, is proportional to the sum of the elongation of the spacer and the shift of the axial positions of the mirrors due to their bending, can then conveniently be written in a compact manner as

$$\Delta \bar{L}_E(p) = \frac{\Delta L_{CS}(p) + 2\Delta L_m(p)}{L_0} \equiv \kappa \frac{p}{E} ,\qquad (S19)$$

where $\kappa$ is a cavity specific dimensionless parameter, defined by the expressions above, that takes all the geometrical and material issues of the cavity spacer and the mirrors of importance into account. For the case when $\Delta L_m(p) \ll \Delta L_{CS}(p)$, $\kappa$ is, for a closed cavity, given by $\beta$. In all other cases, it is larger; for an open cavity, it is 1/3.

Since it is in this case adequate to consider the pressure to be related to the number density of the gas by the ideal gas law, i.e. as $p = kT\rho_n$, the relative strain on the cavity caused by gas with a density of $\rho_n$, $\Delta \bar{L}_E(\rho_n)$, can be written in terms of the gas density, i.e. as

$$\Delta \bar{L}_E(\rho_n) = \rho_n / \rho_c ,\qquad (S20)$$

where $\rho_c$ is given by $E/(\kappa kT)$ [or $E/(\kappa kTN_A)$ if $\rho_c$ is given as a molar density]. As is shown in the footnote below, for a well-designed closed cavity, $\rho_c$ can take values around $1.3 \times 10^9$ mol/m³, while it can be an order of magnitude smaller than this for an open cavity.[S3] Since $\rho_n$ takes a value of 44 mol/m³ under STP conditions, henceforth referred to as $\rho_n^{STP}$, this shows that for a well-designed closed cavity, for which $\beta = 0.03$, $\Delta \bar{L}_E(p)$ is typically around $3 \times 10^{-8}$.

For future use, we will also note that it is possible, by the use of Eq. (6), to express $\rho_n$ in terms of $[2/(3A)](n-1)$. By then defining a dimensionless entity $\varepsilon$ as $2/(3A\rho_c)$, which then also is equal to

$$\varepsilon = \frac{2\kappa kTN_A}{3AE} ,\qquad (S21)$$

henceforth referred to as the cavity deformation parameter, it is possible to express $\Delta \bar{L}_E$ in terms of refractivity as

$$\Delta \bar{L}_E[(n-1)] = \varepsilon \cdot (n-1) .\qquad (S22)$$

For a well-designed closed cavity $\varepsilon$ can take values around $1 \times 10^{-4}$, while it can be an order of magnitude larger for an open cavity.[S4]

Although cavity deformation can be assessed by more detailed descriptions than what is given above, it is sufficient in this case to consider that the instantaneous deformation (of the length) of the cavity is well described by an expression either of the type given by Eq. (S20) or (S22).

*Part 2.2. Effect of light penetration into the mirror coating on the frequency of a longitudinal cavity mode*

It is convenient to describe the penetration of light into the mirror coating in terms of two parts; one is the penetration depth that light of a given frequency will experience when it is reflected by the mirror, while the other is its dispersion, i.e. the phenomenon that the penetration depth has a dependence on frequency, often referred to as mirror dispersion. For this reason, the total penetration depth, $L_{pd}(\nu)$, can formally be decomposed into two parts; *i*) one frequency independent, i.e. the penetration depth for a single mirror at a given frequency, henceforth denoted $L_{pd}^0$, and *ii*) one that takes dispersion into account, denoted $\Delta L_{pd}(\nu)$, i.e. as

$$L_{pd}(\nu) = L_{pd}^0 + \Delta L_{pd}(\nu) .\qquad (S23)$$

***Frequency independent penetration depth:*** For the case when the outermost layer of the antireflection coating has a higher index of reflection than the next layer, $L_{pd}^0$ can be expressed as [12]

$$L_{pd}^0 = \frac{n\lambda_B}{4\bar{n}_m \Delta n_m} ,\qquad (S24)$$

where $\lambda_B$ is the Bragg wavelength, defined as $2\bar{n}_m\Lambda$, $\bar{n}_m$ is the average refractive index of the antireflection coating, which for a well-designed mirror (for which $d_H n_H = d_L n_L$, where $d_H$ and $d_L$ are the thicknesses of the layers with high and low index of reflection, respectively, and $n_H$ and $n_L$ their indices of refraction) is $2n_H n_L/(n_H + n_L)$, $\Lambda$ is the physical length of the double layer, given by $d_H + d_L$, and $\Delta n_m$ is the difference between the two indices of refraction, i.e. $n_H - n_L$, respectively, all evaluated at given frequency, here denoted $\nu_c$. For the case with a cavity with a length of 20 cm, $\bar{L}_{pd}^0$, defined as $L_{pd}^0/nL_0$, will typically take values around $10^{-6}$.[S5]

However, since this penetration depth is frequency independent, it does not contribute significantly to the change of cavity mode frequency that takes place following an evacuation of the cavity. In addition, as is discussed in our accompanying paper,[13] since the penetration depth for the mirror coating considered (with the outermost layer having the highest index of refraction) is proportional to the index of refraction inside the cavity, it does not have any appreciable effect on the accuracy of the technique, in

---

[S1] Other effects, e.g. the alteration of the radius of curvature of the mirrors, produce in general contributions to the change in cavity mode frequency that are smaller than that from the shift of the center positon and have therefore here been neglected.

[S2] Assuming silicon substrates, for which the Poisson's ratio and the shear modulus are 0.28 and 80 GPa respectively, with a thickness of 5 mm mounted on a 200 mm spacer with a bore diameter of 5 mm, the relative deformation from the cavity mirrors, Eq. (S18) shows that $\Delta \bar{L}_m$, defined as $\Delta L_m / L_0$, will then, under STP conditions, be in the order of $10^{-10}$.

[S3] Since $E$ often is in the order of 90 GPa, and the geometrical factor, $\kappa$, can be around 0.03 for a well-designed closed cavity, $\rho_c$ can take values around $1.3 \times 10^9$ mol/m³. For the case with an open cavity, for which $\kappa$ is roughly one order of magnitude larger, $\rho_c$ becomes about one order of magnitude smaller than this.

[S4] Since $\rho_c$ can, for a well-designed closed cavity, take a value of $1.3 \times 10^9$ mol/m³, and since, for $N_2^{STP}$ at 632 nm, $A$ is $4.45 \times 10^{-6}$ m³/mol, this implies that $\varepsilon$ can be estimated to be around $1 \times 10^{-4}$. For a less well-designed cavity, or for an open cavity, $\varepsilon$ will be larger than this; for an open cavity approximately one order of magnitude larger.

[S5] Mirrors can in general be made with various types of coatings. Although these have dissimilar indices of refraction, it is in general found that $L_{pd}^0$ is typically a fraction of $\lambda_B$. For the case with a Ta$_2$O$_5$/SiO$_2$ coating, for which $n_H$ and $n_L$ are around 2.1 and 1.46 (2.04111 and 1.455 at 852 nm),[48] $L_{pd}^0$ is around 0.22 $\lambda_B$, which, for a mirror made for 1.5 µm, corresponds to 0.3 µm. For some other types of coating, t ex Si/SiO$_2$ or TiO$_2$/SiO$_2$, $L_{pd}^0$ can be smaller than this, even < 0.1 $\lambda_B$. Since the relative elongation of the cavity due to a finite penetration depth, $2\bar{L}_{pd}^0$, is given by $2L_{pd}^0/nL_0$, it can be estimated that this entity can, for a cavity with a length of 0.20 m, take a value around $3 \times 10^{-6}$. For other types of coatings, this number can be smaller, even in the high $10^{-7}$ range.

contrast to what previously has been mentioned in the literature.[14] This is not the case for mirrors for which the outermost layer has a lower index of refraction than the next layer.[12] The aforementioned coating configuration, with the outermost layer having the highest index of refraction, is therefore the one preferred for FPC-OR.

***Mirror dispersion:*** It has previously been shown that the frequency-dependent mode spacing of longitudinal cavity modes (i.e. the free-spectral range) of a Fabry-Perot etalon in which there are sources of frequency-dependent phase delays, $\phi(\omega)$, can be expressed as [15-17]

$$\nu_{FSR}(\omega) = \frac{c}{2\{L + c[\partial \phi(\omega)/\partial \omega]\}}, \quad (S25)$$

where $\omega$ is the angular frequency of the light. This shows that the effect of dispersion on the single pass length of the cavity is given by $c\partial \phi/\partial \omega$. In our case, for which $L_{pd}(\nu)$ is defined as the penetration depth of a single mirror, $L_{pd}(\nu)$ is given by half of this, i.e. by $(c/2)\partial \phi/\partial \omega$.

To assess the influence of dispersion of the mirrors (or the penetration depth) on the shift of the frequency of a laser field locked to a mode of a drift-free optical cavity as the cavity is evacuated it is therefore suitable to look at the phase shift that light of a given frequency experiences as it is reflected by the mirror. To do this, it has been found suitable to express the phase shift in terms of a Taylor series around a given center frequency $\omega_c$, *viz.* as

$$\phi(\omega) = \phi(\omega_c) + \left.\frac{\partial \phi(\omega)}{\partial \omega}\right|_{\omega_c} (\omega - \omega_c)$$
$$+ \frac{1}{2} \left.\frac{\partial^2 \phi(\omega)}{\partial \omega^2}\right|_{\omega_c} (\omega - \omega_c)^2 + O\left[(\omega - \omega_c)^3\right], \quad (S26)$$

where $\partial^2 \phi(\omega)/\partial \omega^2\big|_{\omega_c}$, which represents the dispersion, often is denoted $D_2(\omega_c)$.

Inserting this expression for the phase shift into the expression for $L_{pd}(\nu)$ above implies that the latter can be written as

$$L_{pd}(\nu) = \frac{c}{2}\left.\frac{\partial \phi(\omega)}{\partial \omega}\right|_{\omega_c} + \pi c D_2(\omega_c) \cdot (\nu - \nu_c), \quad (S27)$$

where we, in the last term, have expressed angular frequency in normal frequency. The first term represents the frequency independent contribution from the mirror, which above was explicitly given by Eq. (S24), while the last one indicates the effect of dispersion on the penetration depth. Hence, it is this term that represents $\Delta L_{pd}(\nu)$.

It is has become commonplace to assess dispersion of mirrors in terms of the group delay dispersion, $GDD$. The group delay dispersion at the frequency $\nu_c$, $GDD(\nu_c)$, is defined as [16]

$$GDD(\nu_c) = \left.\frac{\partial}{\partial \nu}\frac{1}{\nu_{FSR}(\nu)}\right|_{\nu_c}, \quad (S28)$$

and is often given in units of $(fs)^2$, i.e. $10^{-30}$ s². Inserting the expressions above for the free-spectral range, i.e. Eq. (S25), in this and making use of the series expansion of the phase delay, Eq. (S26), implies that the group delay dispersion can be expressed as [S6]

$$GDD(\nu_c) = 2 \cdot 2\pi \left.\frac{\partial}{\partial \omega}\frac{\partial \phi(\omega)}{\partial \omega}\right|_{\omega_c} = 4\pi D_2(\omega_c). \quad (S29)$$

---

[S6] In the case the mirrors have been made with virtually no group delay dispersion as defined by Eq. (S29), higher order terms in the Taylor expansion of the phase shift, given by Eq. (S27), needs to be included.[45] In those cases, the frequency dependence part of the penetration depth has no linear component with respect to frequency.

This implies that it is possible to write the frequency dependent part of the penetration depth in terms of the $GDD(\nu_c)$ as well as the detuning from $\nu_c$, $\Delta L_{pd}(\nu - \nu_c)$, as

$$\Delta L_{pd}(\nu - \nu_c) = \frac{c}{4} GDD(\nu_c) \cdot (\nu - \nu_c). \quad (S30)$$

It should be noted that in the case the frequency independent penetration depth can be assessed at $\nu_0$, i.e. that $\nu_c = \nu_0$, the $\nu - \nu_c$ entity will represent the detuning from the empty cavity frequency, which is advantageous for analytical purposes. We will therefore, in this work, for simplicity henceforth assume that this is the case.

We will for future convenience also introduce a dimensionless (normalized) mirror penetration dispersion entity, denoted $\Delta \bar{L}_{pd}$, defined as $\Delta L_{pd}/(nL_0)$, and expressed as

$$\Delta \bar{L}_{pd}(\nu - \nu_0) = \varsigma \frac{\nu - \nu_0}{\nu_0}, \quad (S31)$$

where $\varsigma$ is a dimensionless entity given by

$$\varsigma = \frac{c\nu_0}{4nL_0} GDD(\nu_c) = \frac{\nu_{FSR}^0 \nu_0}{2} GDD(\nu_c), \quad (S32)$$

where we in the last step have used the definition of the free spectral range of the cavity.

Mirrors can in general be made with a variety of group delay dispersion. While low-dispersion mirrors can provide $GDD$ values that are only a few tens of fs² (and even close to zero when specifically made for a given frequency),[15-18] ordinary (or high dispersion) mirrors can take values that are several orders of magnitude higher than this, even up to tens of thousands of fs².[19] This implies that for low dispersion mirrors, for which $GDD(\nu_c)$ can be 20 fs², $\varsigma$ will, for the case with a cavity with a length of 30 cm, take a value of $1 \times 10^{-6}$. The corresponding value for high dispersion mirrors [for which $GDD(\nu_c)$ can be as high as 20 000 fs²] is $1 \times 10^{-3}$. Since $(\nu - \nu_0)/\nu_0$ typically is $3 \times 10^{-4}$ under STP conditions, this implies that in the case with no relocking, $\Delta \bar{L}_{pd}$ can range from $3 \times 10^{-10}$ to $3 \times 10^{-7}$.

*Part 2.3. Effect of dispersion of the gas on the frequency of a longitudinal cavity mode*

Also the index of refraction of the gas exhibits dispersion. This implies that we can formally consider $n(\nu)$ to be given by

$$n(\nu) = n(\nu_0) + \left.\frac{\partial n}{\partial \nu}\right|_{\nu_0} (\nu - \nu_0)$$
$$+ \frac{1}{2} \left.\frac{\partial^2 n}{\partial \nu^2}\right|_{\nu_0} (\nu - \nu_0)^2 + O\left[(\nu - \nu_0)^3\right]. \quad (S33)$$

Gases have though in general low dispersion. Since $\partial n/\partial \nu$ is in the order of $6 \times 10^{-21}$ Hz⁻¹ for nitrogen (at 1530 nm),[9] it can be estimated that solely the first order dispersion of the gas addressed can potentially rival other non-linear processes in FCP-OR. Since all successively higher order non-linear terms will be considerably smaller than the previous ones, this show that it is possible to neglect all higher order dispersion terms. It is therefore possible to express the index of refraction of the gas as

$$n(\nu) = n(\nu_0)\left[1 + \Delta \bar{n}(\nu - \nu_0)\right], \quad S(34)$$

where $\Delta \bar{n}(\nu - \nu_0)$ represents the relative shift of the index of refraction due to dispersion, given by

$$\Delta \bar{n}(\nu - \nu_0) = \eta \frac{\nu - \nu_0}{\nu_0}, \quad (S35)$$

where the dimensionless entity $\eta$ is given by

$$\eta = \frac{\nu_0}{n(\nu_0)} \cdot \left.\frac{\partial n}{\partial \nu}\right|_{\nu_0}. \tag{S36}$$

For the case with nitrogen considered above, $\eta$ takes a value of $1.2 \times 10^{-6}$. Moreover, for the case when the frequency of 1.5 µm light is shifted 60 GHz, which corresponds to the shift of a cavity mode in air under STP conditions, this shows that dispersion of the gas can contribute to the assessment of refractivity by a value of $3.6 \times 10^{-10}$.

*Part 2.4. Expression for the frequency shift of a laser locked to a mode of a drift-free optical cavity as the cavity is evacuated*

When the amount of gas in the cavity is rapidly changed from $\rho_i$ to $\rho_f$ (where the subscripts $i$ and $f$ stand for the initial and the final conditions, respectively), the total optical cavity length, $L_{eff}$, is changed from $L_{eff}^i$ to $L_{eff}^f$. Using the expressions from above, it is possible to express these two as

$$\begin{aligned}L_{eff}^i &= n_i L_i + 2L_{pd}^i \\ &= n_i(\nu_0)(1+\Delta\bar{n}_i)(L_0 + \Delta L_E^i) + 2L_{pd}^{0,i} + 2\Delta L_{pd}^i \\ &\approx n_i(\nu_0)L_0\left[1 + \Delta\bar{L}_E^i + \Delta\bar{n}_i + 2\bar{L}_{pd}^0 + 2\Delta\bar{L}_{pd}^i\right] \\ &= n_i(\nu_0)L_0\left[1 + 2\bar{L}_{pd}^0 + \varepsilon(n_i-1) + (2\varsigma+\eta)\frac{\nu_q(n_i)-\nu_0}{\nu_0}\right]\end{aligned} \tag{S37}$$

and

$$\begin{aligned}L_{eff}^f &= n_f L_f + 2L_{pd}^f \\ &= n_f(\nu_0)L_0\left[1 + 2\bar{L}_{pd}^0 + \varepsilon(n_f-1) + (2\varsigma+\eta)\frac{\nu_q(n_f)-\nu_0}{\nu_0}\right],\end{aligned} \tag{S38}$$

where we have neglected the terms that are significantly smaller than $10^{-10}$ of the leading term and made use of the dimensionless entities defined above. Note that we have here included the refractivity of the gas that is in the cell after the evacuation, $n_f - 1$, henceforth referred to as the residual gas refractivity. This implies that the frequency of the $q$:th longitudinal mode of the cavity will shift an amount, $\Delta\nu_l(n_i, n_f)$, that is given by $\nu_q(n_f) - \nu_q(n_i)$, where each frequency, in turn, is given by Eq. (S14) in which the effective optical cavity lengths are given by the expressions above.

It should also be noticed that for the particular type of coating considered (when the outermost layer of the antireflection coating has a higher index of reflection than the following layer), Eq. (S24) shows that the frequency independent penetration depth of the mirrors, $L_{pd}^0$, is proportional to the index of refraction in the cavity. This implies that by defining $\bar{L}_{pd}^0$ as $L_{pd}^0/nL_0$, the normalized penetration depths are independent of the refractivity of the gas in the cavity. This implies particularly that $\bar{L}_{pd}^{0,i} = \bar{L}_{pd}^{0,f}$, whereby they both have been denoted $\bar{L}_{pd}^0$.

For the case when the laser is relocked to a cavity mode whose mode number differ by $\Delta q$, from an initial mode with mode number $q+\Delta q$ to a final mode with mode number $q$, it is possible to express the change in the frequency of the laser field, $\Delta\nu_l(n_i,n_f,\Delta q)$, as $\nu_q(n_f) - \nu_{q+\Delta q}(n_i)$. In this case, $L_{eff}^i$ will be given by

$$L_{eff}^i = n_i(\nu_0)L_0\left[1 + 2\bar{L}_{pd}^0 + \varepsilon(n_i-1) + (2\varsigma+\eta)\frac{\nu_{q+\Delta q}(n_i)-\nu_0}{\nu_0}\right]. \tag{S39}$$

Let us also conclude that when the finite penetration depth of the mirrors is considered, the frequency of the $q$:th longitudinal mode of a cavity in vacuum, $\nu_0$, is given by

$$\nu_0 = \frac{qc}{2L_0(1+2\bar{L}_{pd}^0)}. \tag{S40}$$

By using the expressions derived above, the change in the frequency of the laser field following an evacuation can thereby be written as

$$\begin{aligned}\Delta\nu_l(n_i,n_f,\Delta q) &= \nu_q(n_f) - \nu_{q+\Delta q}(n_i) \\ &= \frac{qc}{2L_{eff}^f} - \frac{(q+\Delta q)c}{2L_{eff}^i} \\ &= \frac{qc}{2L_0}\left\{\frac{1}{n_f\left[1+2\bar{L}_{pd}^0+\varepsilon(n_f-1)+(2\varsigma+\eta)\frac{\nu_q(n_f)-\nu_0}{\nu_0}\right]}\right. \\ &\quad \left. -\left(1+\frac{\Delta q}{q_0}\right)\frac{1}{n_i\left[1+2\bar{L}_{pd}^0+\varepsilon(n_i-1)+(2\varsigma+\eta)\frac{\nu_{q+\Delta q}(n_i)-\nu_0}{\nu_0}\right]}\right\} \\ &= \nu_0\left\{\frac{1}{n_f\left[1+\frac{\varepsilon(n_f-1)}{1+2\bar{L}_{pd}^0}+\frac{2\varsigma+\eta}{1+2\bar{L}_{pd}^0}\frac{\nu_q(n_f)-\nu_0}{\nu_0}\right]}\right. \\ &\quad -\frac{1}{n_i\left[1+\frac{\varepsilon(n_i-1)}{1+2\bar{L}_{pd}^0}+\frac{2\varsigma+\eta}{1+2\bar{L}_{pd}^0}\frac{\nu_{q+\Delta q}(n_i)-\nu_0}{\nu_0}\right]} \\ &\quad \left. -\frac{\Delta q}{q_0}\frac{1}{n_i\left[1+\frac{\varepsilon(n_i-1)}{1+2\bar{L}_{pd}^0}+\frac{2\varsigma+\eta}{1+2\bar{L}_{pd}^0}\frac{\nu_{q+\Delta q}(n_i)-\nu_0}{\nu_0}\right]}\right\}.\end{aligned} \tag{S41}$$

It can be estimated that all non-unity entities in the denominators of the last three terms, i.e. $\varepsilon(n_f-1)$, $\varepsilon(n_i-1)$, $\varsigma[\nu_q(n_f)-\nu_0]/\nu_0$, $\eta[\nu_q(n_f)-\nu_0]/\nu_0$, $\varsigma[\nu_{q+\Delta q}(n_i)-\nu_0]/\nu_0$, $\eta[\nu_{q+\Delta q}(n_i)-\nu_0]/\nu_0$ and $2\bar{L}_{pd}^0$ are $\ll 1$.[S7] This implies that the three terms in Eq. (S41) can be series expanded in terms of these. Such a series expansion can be written as

$$\begin{aligned}\Delta\nu_l(n_i,n_f,\Delta q) &\approx \nu_0\left\{\frac{n_i-n_f}{n_i n_f}(1+\varepsilon) - (2\varsigma+\eta)\frac{\Delta\nu_l}{\nu_0}\right. \\ &\quad \left. -\frac{\Delta q}{q_0}\frac{1}{n_i}\left[1-\varepsilon(n_i-1)-(2\varsigma+\eta)\frac{\Delta\nu_l}{\nu_0}\right]\right\},\end{aligned} \tag{S42}$$

where we have neglected all terms that are expected to be $< 10^{-10}$ with respect to the leading term under STP conditions and assumed that $\nu_q(n_f) - \nu_0$ is so small that $\nu_{q+\Delta q}(n_i) - \nu_0$ can be written as $\Delta\nu_l$.

Note that the frequency dependent penetration depth, represented by $\bar{L}_{pd}^0$, does not affect the shift of the cavity mode noticeably. The

---

[S7] Since, for a well-designed closed cavity under STP conditions, $\varepsilon$ is around $10^{-4}$, $\varepsilon(n_i-1)$ is around $3 \times 10^{-8}$ and hence $\ll 1$. Since $n_f - 1 \ll n_i - 1$, $\varepsilon(n_f-1)$ is considerably smaller than this, thus also $\ll 1$. For an open cavity, the $\varepsilon(n_i-1)$ and $\varepsilon(n_f-1)$ are typically an order of magnitude larger but still $\ll 1$. As was alluded to above, $\Delta\bar{L}_{pd}$, which is strongly dependent on the type of mirror used, is around $10^{-9}$ for a low-dispersion mirror while it can be several orders of magnitude larger for other types. Since $\varsigma[\nu_{q+\Delta q}(n_i)-\nu_0]/\nu_0$ is nothing but $\Delta\bar{L}_{pd}$, $\varsigma[\nu_{q+\Delta q}(n_i)-\nu_0]/\nu_0 \ll 1$. Since $n_f - 1 \ll n_i - 1$, also $\varsigma[\nu_q(n_f)-\nu_0]/\nu_0$ will be $\ll 1$. Since it was estimated above that $\eta \cdot (\nu-\nu_0)/\nu_0$ is about $3.6 \times 10^{-10}$, it should be clear that both $\eta[\nu_{q+\Delta q}(n_i)-\nu_0]/\nu_0$ and $\eta[\nu_q(n_f)-\nu_0]/\nu_0 \ll 1$. $\bar{L}_{pd}^0$, finally, is typically around $10^{-6}$, depending on the type of mirror coating, hence also $\ll 1$.

reason is that it is not affected by the gas evacuation process, wherefore the various entries cancel.

This shows that even though OR can be used to assess changes in the index of refraction of a gas, it is not a technique in which the measured entity, the frequency shift of a cavity mode, which suitably can be measured as the frequency shift of narrow linewidth laser light locked to a given cavity mode as $\Delta \nu_l (n_i, n_f, \Delta q = 0)$, is strictly proportional to the change in refractivity of the gas, $\Delta(n-1)$, defined as $(n_f - 1) - (n_i - 1) = n_f - n_i$; it is affected also by the strain of the cavity and bending of the mirrors when filled with the gas, represented by the $\varepsilon$-parameter, the dispersion of the mirrors as well as of the gas, represented by the $\varsigma$ and the $\eta$ terms, as well as the indices of refraction before and after the evacuation of the cavity, i.e. $n_i$ and $n_f$.

Equation (S42) shows that whereas the elastic elongation of the cavity, represented by $\varepsilon$, affects an assessment of the shift of the laser frequency by an amount that is proportional to the amount of gas in the cavity, represented by the difference in refractivity, $n_i - n_f$, it is influenced by dispersion (from the mirrors as well as the gas) solely in proportion to the relative shift of the laser frequency, $\Delta \nu_l / \nu_0$.

For the case with no relocking, these two entities are identical, whereby the effect of elastic elongation and dispersion will affect an assessment in similar manners; the first set of bracket in Eq. (S42) will basically read $(n_i - n_f)(1 + \varepsilon - 2\varsigma - \eta)/(n_i n_f)$. However, when relocking takes place, the shift of the frequency of the laser is significantly smaller (due to the mode hops), wherefore dispersion plays a relatively smaller role.

As is further discussed in our accompanying paper,[13] while some of these entities affect the accuracy of the instrumentation, i.e. by affecting the linear response, others brings in non-linearities in the response.

Equation (S42) also shows that when the laser field is relocked to a different cavity mode, i.e. when $\Delta q \neq 0$, the measured shift in frequency of the laser field is reduced by an amount that is given by a term that is basically represented by $\Delta q$ times the FSR of the empty cavity in agreement with what is predicted by Eq. (S5) for the simplified description.

*Part 2.5. Expression for the refractivity in terms of the frequency shift of a laser locked to a mode of a drift-free optical cavity as the cavity is evacuated*

To use the shift of the frequency of a laser field, $\Delta \nu_l (n_i, n_f, \Delta q)$, for an assessment of the refractivity in the cavity prior to evacuation, i.e. $n_i - 1$, it is suitable to express Eq. (S42) in terms of the refractivity of the gas prior to, and after, the change of the amount of gas in the cavity, $n_i - 1$ and $n_f - 1$, respectively, *viz.* as

$$\Delta \nu_l (n_i, n_f, \Delta q) = \nu_0 \left\{ \frac{(n_i - 1) - (n_f - 1)}{[1 + (n_i - 1)][1 + (n_f - 1)]} (1 + \varepsilon) \right.$$
$$- (2\varsigma + \eta) \frac{\Delta \nu_l}{\nu_0} \qquad (S43)$$
$$\left. - \frac{\Delta q}{q_0} \frac{1 - \varepsilon(n_i - 1) - (2\varsigma + \eta)\frac{\Delta \nu_l}{\nu_0}}{1 + (n_i - 1)} \right\}.$$

By solving Eq. (S43) for the refractivity of the gas in the cavity prior to its evacuation, e.g. $n_i - 1$, it is possible to obtain an expression for this refractivity in terms of the frequency shift of the cavity mode, $\Delta \nu_l$, any possible change of cavity mode by relocking, $\Delta q$, the parameters for elastic elongation, $\varepsilon$, and those for dispersion, $\varsigma$ and $\eta$, as well as the residual gas after the evacuation, i.e. $n_f - 1$, according to

$$n_i - 1 =$$
$$\frac{\frac{\Delta q}{q_0} + \Upsilon(\varsigma, \eta) \frac{\Delta \nu_l}{\nu_0} + 2\varsigma \frac{\Delta \nu_l}{\nu_0} \frac{\Delta q}{q_0} + (n_f - 1)\left[1 + \varepsilon + \Upsilon(\varsigma, \eta) \frac{\Delta \nu_l}{\nu_0} + \frac{\Delta q}{q_0}\right]}{1 + \left(1 + \frac{\Delta q}{q_0}\right)\varepsilon - \Upsilon(\varsigma, \eta) \frac{\Delta \nu_l}{\nu_0} - (n_f - 1)\left[\Upsilon(\varsigma, \eta) \frac{\Delta \nu_l}{\nu_0} - \varepsilon \frac{\Delta q}{q_0}\right]}$$
. (S44)

where we have neglected terms that clearly are smaller than the leading one by $10^{-10}$ (or more) under STP conditions and where $\Upsilon(\varsigma, \eta)$ is given by $1 + 2\varsigma + \eta$.[S8] It also can be noticed that, if necessary, $n_f - 1$ can be expressed in terms of the residual gas density in the cavity, $\rho_{n,f}$, by the use of Eq. (6).

Again, since all non-unity dimensionless entities are significantly smaller than unity, it is possible to series expand this expression in terms of these. Although such a series expansion is rather lengthy and therefore not explicitly given here, it can be scrutinized for its behavior for a variety of situations. We will do so here and specifically distinguish between the cases when the laser field is locked to the cavity mode during the entire evaluation process (no relocking, i.e. when $\Delta q = 0$) and when it is relocked, to a different cavity mode (when $\Delta q \neq 0$).

*Part 2.6. Simplified expressions for the refractivity in terms of the shift of the frequency of a laser locked to a mode of a drift-free optical cavity as the cavity is evacuated without relocking*

Series expanding Eq. (S44) and again neglecting terms that presumably are smaller than the leading one by $10^{-10}$ (or more) under STP conditions provides, for the case when the laser is kept locked to the same mode during evacuation of the cavity, i.e. when $\Delta q = 0$, an expression for the refractivity of the gas prior to evacuation in terms of the frequency of the cavity mode, $\Delta \nu_l$, the deformation of the cavity, represented by the $\varepsilon$-parameter, and the dispersion of the mirrors, by the $\varsigma$ and $\eta$-parameters that reads

$$n_i - 1 = \frac{\Delta \nu_l}{\nu_0} \Omega(\varepsilon, \varsigma, \eta) \left[1 + \Omega(\varepsilon, \varsigma, \eta) \frac{\Delta \nu_l}{\nu_0} + \left(\frac{\Delta \nu_l}{\nu_0}\right)^2\right]$$
$$+ (n_f - 1)\left[1 + 2\Omega(\varepsilon, \varsigma, \eta) \frac{\Delta \nu_l}{\nu_0} + 3\left(\frac{\Delta \nu_l}{\nu_0}\right)^2\right], \qquad (S45)$$

where we have used $\Omega(\varepsilon, \varsigma, \eta)$ as a shorthand notation $1 - \varepsilon + 2\varsigma + \eta + \varepsilon^2$.

Although it is feasible to assume that it is possible to evacuate the cavity to such an extent that the refractivity in the evacuated cell can be neglected, we have here, to illustrate the dependence of $n_i - 1$ on $n_f - 1$, kept the leading terms of this dependence.[S9]

This shows that the refractivity is affected by the distortion of the cavity as well as the mirrors, i.e. $\varepsilon$, $\varsigma$, and $\eta$, and that it is non-linear in the relative shift of a mode of a cavity that follows an evacuation, i.e. $(\Delta \nu_l / \nu_0)$. Since, for the case when gases at STP

---

[S8] The $(2\varsigma + \eta)(\Delta \nu_l / \nu_0)$ term in Eq. (S43) can often be neglected. It can first be concluded that since $\eta$ typically is around $10^{-6}$ while $\Delta \nu_l / \nu_0$ is typically smaller than $\nu_0^{FSR} / \nu_0$, whenever relocking take place (further alluded to elsewhere in the text) the $\eta(\Delta \nu_l / \nu_0)$ entity is clearly $< 10^{-10}$. It is also possible to conclude that $2\varsigma(\Delta \nu_l / \nu_0)$ will be below $10^{-10}$ whenever $\varsigma < 2 \times 10^{-5}$ (for the case with a FSR of 500 MHz). Since it was shown above that $\varsigma$ can be in the order of $10^{-6}$ for low dispersion mirrors, this implies that also the $2\varsigma(\Delta \nu_l / \nu_0)$ term in Eq. (S43) often can be neglected. However, to allow for other types of mirrors, the $2\varsigma(\Delta \nu_l / \nu_0)$ entity has been kept in its most prominent position in Eq. (S44).

[S9] A pressure of 1 mTorr produces in general a refractivity of a few times $10^{-10}$ (4 × $10^{-10}$ for $N_2^{STP}$ at 1.5 μm). Since $\Delta \nu_l / \nu_0$ is in the order of 3 × $10^{-4}$, all higher order terms of the $n_f - 1$ factor can be neglected whenever the residual pressure is below this.

conditions are assessed, the relative shift of a mode of a cavity that follows an evacuation, i.e. $(\Delta \nu_l / \nu_0)$, can take values up to around a few times $10^{-4}$ ($3 \times 10^{-4}$ for $N_2^{STP}$ at 1.5 µm), the refractivity will be affected by two high-order contributions of the relative shift of a mode of a cavity that follows an evacuation. This shows that the non-linearities in the system play a successively larger role the larger gas densities that are assessed and that they take place on the $3 \times 10^{-4}$ level under STP conditions but to a lesser extent for more diluted gases. The effects of distortion of the cavity and the dispersion effects, on the other hand, affect the measurements already for low density assessments. As is further alluded to in our accompanying paper,[13] this shows that if the highest accuracies are to be obtained, it is of highest importance to construct a system with a minimum of cavity distortion and to use low-dispersion mirrors.

*Part 2.7. Simplified expressions for the refractivity in terms of the frequency shift of a laser locked to a mode of a drift-free optical cavity as the cavity is evacuated in the presence of relocking*

It is possible to series expand Eq. (S44) also for the case when the laser is relocked to a different cavity mode. When relocking takes place, however, it is customary to relock to a mode so that $\Delta \nu_l / \nu_0$ is small, *viz.* in the order of (or below) $\nu_{FSR}^0 / \nu_0$, which, for the case with a cavity with a length of 25 cm, for which $\nu_{FSR}^0$ is 600 MHz, is around $3 \times 10^{-6}$ at 1.5 µm, while $\Delta q / q_0$ often is in the order of $n_i - 1$, which for STP conditions can take values up to around a few times $10^{-4}$ ($3 \times 10^{-4}$ for $N_2^{STP}$ at 1.5 µm). This implies that the various terms Eq. (S44) have slightly different relative values for the case with relocking than for the case with no relocking. Based on this, and still under the condition that the cavity is evacuated to such an extent that $(n_f - 1)$ is significantly smaller than both $\Delta \nu_l / \nu_0$ and $\Delta q / q_0$, the series expansion of Eq. (S44) can be written as

$$n_i - 1 = \frac{\Delta q}{q} \left[ 1 - \varepsilon \left( 1 + \frac{\Delta q}{q} \right) + \varepsilon^2 \right]$$
$$+ \frac{\Delta \nu}{\nu_0} \left( 1 + \frac{\Delta q}{q} \right) \left\{ \Omega(\varepsilon, \varsigma, \eta) - 2 \left[ \varepsilon(1 - 2\varepsilon) - \varsigma \right] \frac{\Delta q}{q} \right\}$$
$$+ \left( \frac{\Delta \nu}{\nu_0} \right)^2 \left( 1 + \frac{\Delta q}{q} \right) \left[ 2\Omega(\varepsilon, \varsigma, \eta) - 1 \right]$$
$$+ (n_f - 1) \left( 1 + \frac{\Delta q}{q} \right) \left[ 1 - 2\varepsilon \frac{\Delta q}{q} + 2\Omega(\varepsilon, \varsigma, \eta) \frac{\Delta \nu_l}{\nu_0} \right].$$ (S46)

This shows, first of all, in agreement with the first order expression given above, Eq. (5) that the leading term constitutes a sum of $\Delta q / q_0$ and $\Delta \nu_l / \nu_0$. Since $\Delta \nu_l / \nu_0$ presumably is considerably smaller than $\Delta q / q_0$, it also illustrates that the leading term, i.e. the one proportional to $\Delta q / q_0$, is affected by cavity deformation but not dispersion effects. The reason for this is that while cavity deformation affects all terms, dispersion effects only affects the terms that are proportional to $\Delta \nu_l / \nu_0$. This shows that dispersion plays a significantly smaller role for the case when relocking is used than when the laser is not relocked.

In conclusion, both the last two expressions, Eqs. (S45) and (S46), show clearly that refractivity is not fully linear with the relative change in frequency of the cavity mode (or alternatively the frequency of the laser light). This implies that the density of gas molecules in a cavity is not only non-linearly dependent on the refractivity according to Eq. (16) or (17) the refractivity is additionally non-linear with respect to the relative change in frequency of the cavity mode (alternatively the frequency of the laser light) according to either of the Eqs. (S45) and (S46).

**Part 3 – Derivation of an expression for the density of a real gas in terms of refractivity**

For a real gas subjected to two- and three-body interactions, the index of refraction can be related to the density of the gas by the virial expansion of the Lorentz-Lorenz equation (also known as the Clausius–Mossotti relation), which can be written as

$$\frac{n^2 - 1}{n^2 + 2} = A_R \rho_n + B_R \rho_n^2 + C_R \rho_n^3 + \ldots,$$ (S47)

where $A_R$, $B_R$, and $C_R$ are so called virial coefficients.[20] In this case, by comparison with the Eqs. (S7) and (S10), $A_R$ can be identified as the molar polarizability (or molar refractivity) of an ideal gas, while the higher order terms, i.e. those containing the $B_R$ and $C_R$ coefficients (known as the dielectric virial coefficients),[20] represent the influence of mutual interactions of the molecules on their polarizability.[20-26]

As was alluded to in part 1 of the supplementary material, when the gas is not considered as thin, it is not appropriate to consider the left hand side of the Lorentz-Lorenz equation to be well represented by the first (linear) term in the series expansion of it around $(n - 1)$, as was done when deriving Eq. (S9) by use of Eq. (S8). It is therefore of importance to use Eq. (S47) with as little modifications as possible.

However, since Eq. (S47) is written in an implicit form, *viz.* the left hand side consists of a ratio of two second order polynomials in the index of refraction, while, in the presence of multi-body interactions, the right hand side consists of a higher order polynomial in the density of molecules, it is not trivial to derive an expression that relates $n - 1$ to $\rho_n$.

One way of proceeding though is to assume that the refractivity can be written as a convergent series in terms of increasing power of the density, i.e. as

$$n - 1 = A_m \rho_n + B_m \rho_n^2 + C_m \rho_n^3 + \ldots.$$ (S48)

Inserting this expression into Eq. (S47) and equalizing terms with the same power of gas density shows that the $A_m$, $B_m$, and $C_m$ coefficients can be expressed in terms of the $A_R$, $B_R$, and $C_R$ coefficients, *viz.* as $3A_R / 2$, $3B_R / 2 + 3A_R^2 / 8$, and $3C_R / 2 + 3A_R B_R / 4 + 15A_R^3 / 16$, respectively. Mathematically, this can be seen as an expansion of Eq. (S47) around $n - 1$.

When gas density is to be assessed, this expression is still not written in its most useful form; it would be more useful if it could be written as $\rho_n = f(n-1)$. Again, since Eq. (S48) can be seen as a convergent series of successively higher order contributions that contribute less and less to the refractivity, it is possible to "invert" it, i.e. to express the density in a convergent series of terms of the refractivity, *viz.* as

$$\rho_n = A_\rho (n-1) + B_\rho (n-1)^2 + C_\rho (n-1)^3 + \ldots,$$ (S49)

where the various expansion coefficients, $A_\rho$, $B_\rho$, and $C_\rho$ coefficients can be given in terms of the $A_m$, $B_m$, and $C_m$ coefficients, *viz.* as $1/A_m$, $-B_m / A_m^3$, and $(2B_m^2 - A_m C_m)/A_m^5$, respectively. Inserting the expressions for the $A_m$, $B_m$, and $C_m$ coefficients from above provides expressions for the $A_\rho$, $B_\rho$, and $C_\rho$ coefficients in terms of the $A_R$, $B_R$, and $C_R$ virial coefficients that are given Table 1.

However, although being formally correct, it is often more convenient to write this expression as

$$\rho_n = \frac{2}{3A_R}(n-1)\left[1 + \tilde{B}_\rho (n-1) + \tilde{C}_\rho (n-1)^2 + \ldots \right],$$ (S50)

where the leading term, which provides the linear $(n-1)$-dependence, has been extracted and its coefficient, $A_\rho$, explicitly expressed, $2/(3A_R)$. The normalized higher order coefficients, $\tilde{B}_\rho$ and $\tilde{C}_\rho$, which thus are given by the ratios $B_\rho / A_\rho$, and $C_\rho / A_\rho$, respectively, are also given in Table 1, likewise expressed in terms of the $A_R$, $B_R$, and $C_R$ coefficients.

## Part 4 – Assessment of the density of molecules by measurement of the frequency shift of a laser field locked to a mode of a drift-free optical cavity as the cavity is evacuated

*Part 4.1. Expression for the density of molecules by measurement of the frequency shift of a laser field locked to a mode of a drift-free optical cavity as the cavity is evacuated with no relocking*

By inserting Eq. (13) [or (S45)] for the assessment of refractivity in terms of the shift of the laser frequency (without relocking) as the cavity is evacuated, $\Delta \nu_l$, into Eq. (16) [or (S49)] for the density, $\rho_n$, in terms of refractivity, we can obtain an expression for the density of gas in the cavity prior to the evacuation, $\rho_{n,i}$, expressed in terms of the shift of the laser frequency that can be expressed as

$$\rho_{n,i} = \rho_{n,f} + A_{\Delta \nu} \frac{\Delta \nu_l}{\nu_0} + B_{\Delta \nu}\left(\frac{\Delta \nu_l}{\nu_0}\right)^2 + C_{\Delta \nu}\left(\frac{\Delta \nu_l}{\nu_0}\right)^3 + \ldots \quad \text{(S51)}$$

where $\rho_{n,f}$, which is given by $2(n_f - 1)/(3A_R)$, represents the residual density of the gas after the evacuation, henceforth referred to as the residual gas density. The various expansion coefficients, $A_{\Delta \nu}$, $B_{\Delta \nu}$, and $C_{\Delta \nu}$, are given in terms of the $A_\rho$, $B_\rho$, and $C_\rho$ coefficients as well as the cavity deformation parameter, $\varepsilon$, and the dispersion factors, $\varsigma$ and $\eta$, according to Table S1 where we have neglected the terms that presumably (i.e. under conventional conditions) are smaller than $10^{-10}$ of the leading term. In addition, by using the expressions in Table 1, which relate the $A_\rho$, $B_\rho$, and $C_\rho$ coefficients to the virial $A_R$, $B_R$, and $C_R$ coefficients, it is possible to express also the $A_{\Delta \nu}$, $B_{\Delta \nu}$, and $C_{\Delta \nu}$ coefficients in terms of the virial coefficients. Also these expressions are given in Table S1. Again, for simplicity, we have neglected terms that presumably are smaller than $10^{-10}$ of the leading term.

Table S1. Explicit expressions for the various coefficients in Eq. (S51) for the assessment of gas density for the case with no relocking in terms of either the $A_\rho$, $B_\rho$, and $C_\rho$ coefficients or the $A_R$, $B_R$, and $C_R$ virial coefficients, as well as the cavity deformation parameter, $\varepsilon$, and the dispersion factors, $\varsigma$ and $\eta$.*

| Coefficient | Relation to the $A_\rho$, $B_\rho$, and $C_\rho$ coefficients or the $A_R$, $B_R$, and $C_R$ virial coefficients, as well as the cavity deformation parameter, $\varepsilon$, and the dispersion factors, $\varsigma$ and $\eta$. |
|---|---|
| $A_{\Delta \nu}$ | $\Omega(\varepsilon,\varsigma,\eta)A_\rho + 2(n_f - 1)(A_\rho + B_\rho)$ $= \frac{2}{3A_R}\left[\Omega(\varepsilon,\varsigma,\eta) + \frac{5}{3}(n_f - 1)\left(1 - \frac{4}{5}\frac{B_R}{A_R^2}\right)\right]$ |
| $B_{\Delta \nu}$ | $(1 - 2\varepsilon + 4\varsigma + 2\eta)(A_\rho + B_\rho) + 3(n_f - 1)(A_\rho + 2B_\rho + C_\rho)$ $= \frac{5}{9A_R}\left\{[2\Omega(\varepsilon,\varsigma,\eta) - 1]\left(1 - \frac{4}{5}\frac{B_R}{A_R^2}\right)\right.$ $\left. + (n_f - 1)\frac{8}{5}\left(1 - \frac{C_R}{A_R^3} - \frac{5}{2}\frac{B_R}{A_R^2} + 2\frac{B_R^2}{A_R^4}\right)\right\}$ |
| $C_{\Delta \nu}$ | $A_\rho + 2B_\rho + C_\rho$ $= \frac{8}{27A_R}\left(1 - \frac{C_R}{A_R^3} - \frac{5}{2}\frac{B_R}{A_R^2} + 2\frac{B_R^2}{A_R^4}\right)$ |

* For estimates of the various entities, see the footnote of Table 1.

However, although being formally correct, it is often more convenient to write Eq. (S51) as

$$\rho_{n,i} = \rho_{n,f} + \frac{2}{3A_R}\tilde{\chi}_{\Delta \nu}\frac{\Delta \nu_l}{\nu_0}\left[1 + \tilde{B}_{\Delta \nu}\frac{\Delta \nu_l}{\nu_0} + \tilde{C}_{\Delta \nu}\left(\frac{\Delta \nu_l}{\nu_0}\right)^2 + \ldots\right], \quad \text{(S52)}$$

where the leading term, which provides the linear $\Delta \nu / \nu_0$-dependence, has been extracted from the part of the expressions that has a $\Delta \nu / \nu_0$-dependence, and where $A_{\Delta \nu}$ has been expressed in term of a product of its first order expression, $2/(3A_R)$, and a dimensionless factor, denoted $\tilde{\chi}_{\Delta \nu}$, explicitly given in Table 2, that is close to unity but includes contributions from the deformation of the cavity, $\varepsilon$, the dispersion of the mirrors and the gas, $\varsigma$ and $\eta$, as well as the influence of the residual gas refractivity, i.e. $n_f - 1$. The normalized higher order coefficients, $\tilde{B}_{\Delta \nu}$ and $\tilde{C}_{\Delta \nu}$, which thus are given by the ratios $B_{\Delta \nu}/A_{\Delta \nu}$ and $C_{\Delta \nu}/A_{\Delta \nu}$, respectively, are also explicitly given in Table 2 expressed in terms of the $A_R$, $B_R$, and $C_R$ coefficients. Again, terms that presumably are smaller than $10^{-10}$ of the leading term have been neglected.

Since Table 2 shows that $\tilde{B}_{\Delta \nu}$ is around unity (its leading term is 5/6) while $\tilde{C}_{\Delta \nu}$ is roughly an order of magnitude smaller than this [its leading terms is $(4/9)(1 - C_R/A_R^3)$ where, for $N_2^{STP}$ at 632 nm, $C_R/A_R^3$ is 1.2], it can be concluded that the first non-linear term in Eq. (S52) (the one proportional to $\tilde{B}_{\Delta \nu}$) needs to be included whenever the density of the gas is to be assessed with an accuracy that is in the order of $\Delta \nu_l / \nu_0$ or better.[S10] Since, according to Eq. (3) this roughly represents the refractivity of the sample, i.e. $n - 1$, which, under STP conditions, typically is around $3 \times 10^{-4}$, this implies that the first of the higher order terms needs to be taken into account whenever the density is to be assessed under STP conditions with an accuracy of $3 \times 10^{-4}$ or better.

The second non-linear term (proportional to $\tilde{C}_{\Delta \nu}$) needs to be included only whenever the density of the gas is to be assessed with an accuracy that is in the order of $(\Delta \nu_l / \nu_0)^2 / 10$, which, under STP conditions, for $N_2^{STP}$, amounts to $1 \times 10^{-8}$. Since this term is significantly smaller than the first non-linear term, it can be concluded that it plays an almost insignificant role in the assessment of the density under STP conditions. When larger gas densities are to be assessed though, its role can be more pronounced.

Moreover, it is often possible to evacuate the cavity to such low pressures that the influence of the residual gas, by an assessment of the refractivity, in practice can be neglected.[S11] This implies that it is, under many types of conditions, appropriate to assess the gas density by OR by use of the expression

$$\rho_{n,i} = \rho_{n,f} + \frac{2}{3A_R}\Omega(\varepsilon,\varsigma,\eta)\frac{\Delta \nu_l}{\nu_0}\left(1 + \tilde{B}_{\Delta \nu}\frac{\Delta \nu_l}{\nu_0}\right). \quad \text{(S53)}$$

*Part 4.2. Expression for the density of molecules by measurement of the frequency shift of a laser field locked to a mode of a drift-free optical cavity as the cavity is evacuated in the presence of relocking*

By inserting Eq. (14) [or (S46)], for the assessment of refractivity in terms of the shift of the laser frequency in the presence of relocking as the cavity is evacuated, into Eq. (16) [or (S49)] for the density in terms of refractivity, it is possible to obtain an expression the density of gas in the cavity prior to the evacuation, $\rho_{n,i}$, expressed in terms of both the number of modes by which the laser field is shifted during

---

[S10] It is worth to note that the non-linear term in the expression for $\rho_n$ with respect to $\Delta \nu_l / \nu_0$, as given by the $\tilde{B}_{\Delta \nu}$-term in Eq. (S52), is significantly larger than that for $\rho_n$ with respect to $n - 1$, as given by the $\tilde{B}_\rho$-term in Eq. (S50). Inserting typical numbers for $N_2$ shows that they differ by a factor of 4; $0.81(\Delta \nu_l / \nu_0)$ for the non-linear term in Eq. (S52) versus $-0.19(n - 1)$ for that in Eq. (S50), respectively. This difference originates from the non-linearity of $n - 1$ with respect to $\Delta \nu / \nu_0$, which was given by Eq. (S46).

[S11] It suffices to evacuate the cavity down to 1 mTorr to produce a final refractivity, $n_f - 1$, of a few times $10^{-10}$ ($4 \times 10^{-10}$ for $N_2^{STP}$ at 1.5 μm).

the relocking process, $\Delta q$, and the frequency shift of the laser light, $\Delta \nu_l$, when the cavity is evacuated that can be expressed as

$$\rho_{n,i} = \rho_{n,f} + A_-^{\Delta q} \frac{\Delta q}{q_0} + A_{\Delta \nu}^- \frac{\Delta \nu_l}{\nu_0}$$
$$+ B_-^{(\Delta q)^2} \left(\frac{\Delta q}{q_0}\right)^2 + B_{\Delta \nu}^{\Delta q} \frac{\Delta \nu_l}{\nu_0} \frac{\Delta q}{q_0} + B_{(\Delta \nu)^2}^- \left(\frac{\Delta \nu_l}{\nu_0}\right)^2$$
$$+ C_-^{(\Delta q)^3} \left(\frac{\Delta q}{q_0}\right)^3 + C_{\Delta \nu}^{(\Delta q)^2} \left(\frac{\Delta q}{q_0}\right)^2 \frac{\Delta \nu_l}{\nu_0} \quad (S54)$$
$$+ C_{(\Delta \nu)^2}^{\Delta q} \frac{\Delta q}{q_0} \left(\frac{\Delta \nu_l}{\nu_0}\right)^2 + \ldots,$$

where we have organized the result in terms that are of the form $(\Delta \nu_l / \nu_0)^j (\Delta q / q)^i$ for which $i + j \leq 3$ (but neglected the $(\Delta \nu_l / \nu_0)^3$ term due to its smallness). The various coefficients, $A_-^{\Delta q}$, $A_{\Delta \nu}^-$, etc., were first calculated in terms of the $A_\rho$, $B_\rho$, and $C_\rho$ coefficients (not explicitly given here). Table 1 was then used to relate these to the $A_R$, $B_R$, and $C_R$ virial coefficients. The various coefficients are explicitly given in Table S2. To simplify the expressions, we have again neglected terms that are supposedly $10^{-10}$ smaller than the leading term, which in practice has implied that we have included in the $A$-coefficients the contributions from the elongation and dispersion represented by terms including $\varepsilon$, $\varsigma$, $\eta$, $\varepsilon^2$, as well as $n_f - 1$ (although we have neglected terms containing products of $n_f - 1$ and $\varepsilon^2$), in the $B$-coefficients the terms that have contributions from either of $\varepsilon$, $\varsigma$, $\eta$, and $n_f - 1$, while we in the $C$-coefficients have not included any contribution from these entities.

As above, it also possible to write this expression in a normalized manner, i.e. as

$$\rho_{n,i} = \rho_{n,f} + \frac{2}{3A_R} \tilde{\chi}_{\Delta q} \left\{ \frac{\Delta q}{q_0} \left[ 1 + \tilde{B}_-^{(\Delta q)^2} \frac{\Delta q}{q_0} + \tilde{C}_-^{(\Delta q)^3} \left(\frac{\Delta q}{q_0}\right)^2 \right] \right.$$
$$+ \frac{\Delta \nu_l}{\nu_0} \left[ \tilde{A}_{\Delta \nu}^- + \tilde{B}_{\Delta \nu}^{\Delta q} \frac{\Delta q}{q_0} + \tilde{C}_{\Delta \nu}^{(\Delta q)^2} \left(\frac{\Delta q}{q_0}\right)^2 \right] \quad (S55)$$
$$\left. + \left(\frac{\Delta \nu_l}{\nu_0}\right)^2 \left[ \tilde{B}_{(\Delta \nu)^2}^- + \tilde{C}_{(\Delta \nu)^2}^{\Delta q} \frac{\Delta q}{q_0} \right] \right\},$$

where the leading term, which provides the linear $\Delta q / q_0$-dependence, $[2/(3A_R)]\tilde{\chi}_{\Delta q}$, where the $\tilde{\chi}_{\Delta q}$ factor is a dimensionless factor that is close to unity but takes the influences from the cavity deformation parameter, $\varepsilon$, as well as the influence of the refractivity of the residual gas, i.e. $n_f - 1$, into account, has been extracted. This factor, together with the normalized higher order coefficients $\tilde{B}_-^{(\Delta q)^2}$, $\tilde{C}_-^{(\Delta q)^3}$, $\tilde{A}_{\Delta \nu}^-$, $\tilde{B}_{\Delta \nu}^{\Delta q}$, $\tilde{C}_{\Delta \nu}^{(\Delta q)^2}$, $\tilde{B}_{(\Delta \nu)^2}^-$, and $\tilde{C}_{(\Delta \nu)^2}^{\Delta q}$ are given formally in Table S3 and explicitly in Table 3.

Table S2. Explicit expressions for the various coefficients used for assessment of gas density in terms of the shift of the frequency of a laser locked to a mode of a drift-free optical cavity as the cavity is evacuated in the presence of relocking given by Eq. (S54).*

| Coefficient | Relation to the cavity deformation parameter, $\varepsilon$ as well as the $A_R$, $B_R$, and $C_R$ virial coefficients. |
|---|---|
| $A_-^{\Delta q}$ | $\frac{2}{3A_R}(1 - \varepsilon + \varepsilon^2)$ $+ (n_f - 1)\frac{4}{9A_R}\left[\left(1 - \frac{5}{2}\varepsilon\right) - 2\frac{B_R}{A_R^2}(1 - \varepsilon)\right]$ |
| $A_{\Delta \nu}^-$ | $\frac{2}{3A_R}\Omega(\varepsilon, \varsigma, \eta)$ $+ (n_f - 1)\frac{10}{9A_R}\Omega(\varepsilon, \varsigma, \eta)\left[1 - \frac{4}{5}\frac{B_R}{A_R^2}\right]$ |
| $B_-^{(\Delta q)^2}$ | $\frac{-1}{9A_R}\left[(1 + 4\varepsilon) + 4(1 - 2\varepsilon)\frac{B_R}{A_R^2}\right]$ $- (n_f - 1)\frac{2}{3A_R}\left[\left(1 - \frac{2}{3}\varepsilon\right) + \frac{2}{3}(1 - 6\varepsilon)\frac{B_R}{A_R^2}\right.$ $\left. - \frac{8}{3}(1 - 2\varepsilon)\frac{B_R^2}{A_R^4} + \frac{4}{3}(1 - 2\varepsilon)\frac{C_R}{A_R^3}\right]$ |
| $B_{\Delta \nu}^{\Delta q}$ | $\frac{4}{9A_R}\left[\left(1 - \frac{7}{2}\varepsilon + 2\varsigma + \eta\right) - 2(1 - 2\varepsilon + 2\varsigma + \eta)\frac{B_R}{A_R^2}\right]$ $- (n_f - 1)\frac{4}{9A_R}\left[(1 + 3\varepsilon + 2\varsigma + \eta) + 6\left(1 - \frac{8}{3}\varepsilon + 2\varsigma + \eta\right)\frac{B_R}{A_R^2}\right.$ $\left. - 8(1 - 2\varepsilon + 2\varsigma + \eta)\frac{B_R^2}{A_R^4} + 4(1 - 2\varepsilon + 2\varsigma + \eta)\frac{C_R}{A_R^3}\right]$ |
| $B_{(\Delta \nu)^2}^-$ | $\frac{5}{9A_R}[2\Omega(\varepsilon, \varsigma, \eta) - 1]\left(1 - \frac{4}{5}\frac{B_R}{A_R^2}\right)$ $+ (n_f - 1)\frac{8}{9A_R}[2\Omega(\varepsilon, \varsigma, \eta) - 1]\left[1 - \frac{5}{2}\frac{B_R}{A_R^2} + 2\frac{B_R^2}{A_R^4} - \frac{C_R}{A_R^3}\right]$ |
| $C_-^{(\Delta q)^3}$ | $\frac{-4}{27A_R}\left[1 - \frac{B_R}{A_R^2} - 4\frac{C_R}{A_R^3} + 2\frac{B_R^2}{A_R^4}\right]$ |
| $C_{\Delta \nu}^{(\Delta q)^2}$ | $-\frac{2}{3A_R}\left[1 + \frac{2}{3}\frac{B_R}{A_R^2} - \frac{8}{3}\frac{B_R^2}{A_R^4} + \frac{4}{3}\frac{C_R}{A_R^3}\right]$ |
| $C_{(\Delta \nu)^2}^{\Delta q}$ | $\frac{-2}{9A_R}\left[1 + 6\frac{B_R}{A_R^2} - 8\frac{B_R^2}{A_R^4} + 4\frac{C_R}{A_R^3}\right]$ |

* For estimates of the various entities, see the footnote of Table 1.

Table S3. Explicit expressions for the various coefficients used for assessment of gas density in terms of the shift of the frequency of a laser locked to a mode of a drift-free optical cavity as the cavity is evacuated in the presence of relocking given by Eq. (S55).*

| Coefficient | Relation to the other parameters and coefficients. |
|---|---|
| $\tilde{B}_{-}^{(\Delta q)^2}$ | $\dfrac{B_{-}^{(\Delta q)^2}}{A_{-}^{\Delta q}}$ |
| $\tilde{C}_{-}^{(\Delta q)^3}$ | $\dfrac{C_{-}^{(\Delta q)^3}}{A_{-}^{\Delta q}}$ |
| $\tilde{A}_{\Delta\nu}^{-}$ | $\dfrac{A_{\Delta\nu}^{-}}{A_{-}^{\Delta q}}$ |
| $\tilde{B}_{\Delta\nu}^{\Delta q}$ | $\dfrac{B_{\Delta\nu}^{\Delta q}}{A_{-}^{\Delta q}}$ |
| $\tilde{C}_{\Delta\nu}^{(\Delta q)^2}$ | $\dfrac{C_{\Delta\nu}^{(\Delta q)^2}}{A_{-}^{\Delta q}}$ |
| $\tilde{B}_{(\Delta\nu)^2}^{-}$ | $\dfrac{B_{(\Delta\nu)^2}^{-}}{A_{-}^{\Delta q}}$ |
| $\tilde{C}_{(\Delta\nu)^3}^{-}$ | $\dfrac{C_{(\Delta\nu)^3}^{-}}{A_{-}^{\Delta q}}$ |

\* Expressed in parts in terms of the Tables S2 and 1.